\documentclass{ws-ijmpe}
\usepackage[super,compress]{cite}
\usepackage{hyperref}
\usepackage{upgreek}
\usepackage{xspace}
\usepackage{color}
\usepackage[utf8]{inputenc}

\begin{document}

\markboth{B. D\"{o}nigus}{Light nuclei in the hadron resonance gas}

\catchline{}{}{}{}{}

\title{Light nuclei in the hadron resonance gas}

\author{Benjamin D\"{o}nigus}

\address{Institute for Nuclear Physics, Goethe-University Frankfurt, Max-von-Laue-Str. 1\\
Frankfurt, 60438, Hessen,
Germany\\
benjamin.doenigus@cern.ch}

\maketitle

\begin{history}
\received{13 December 2019}
\revised{9 March 2020}
\accepted{11 March 2020}
\end{history}

\begin{abstract}
The description of the production of light (anti-)(hyper-)nuclei in a hadron resonance gas, or statistical-thermal model approach, has proven to be rather successful, despite the fact that the binding energies of these compound objects are small compared to the emitting fireball temperature. We summarise some recent developments and findings in this approach.   
\end{abstract}

\keywords{Hadron gas; Antinuclei; Hypernuclei.}

\ccode{PACS numbers:21.80.+a, 71.10.Ca, 03.75.Ss, 05.30.-d}

\section{Introduction}
Hadron yields measured in heavy-ion collisions at various energies are known to be described surprisingly well by the thermal model~\cite{Cleymans:1992zc,BraunMunzinger:1994xr,BraunMunzinger:1995bp,BraunMunzinger:1996mq,BraunMunzinger:1998cg,Becattini:2000jw}, which  in the simplest case represents a non-interacting gas of known hadrons and resonances in the grand canonical ensemble~(see, e.g., Ref.~\cite{Braun-Munzinger:2015hba} for an overview).
This concept has also been for many years applied to production yields of light nuclei~\cite{Mekjian:1977ei,Gosset:1988na,Mekjian:1978us,Siemens:1979dz,Stoecker:1981za,Hahn:1986mb,Csernai:1986qf}, and, even more surprisingly, a very good description of the various light (anti-)(hyper-)nuclei yields measured in heavy-ion collisions is obtained~\cite{BraunMunzinger:1994iq,Andronic:2010qu,Steinheimer:2012tb,Adam:2015vda,Adam:2015yta,Anticic:2016ckv}. 

Interestingly, many people connect the production of light nuclei in high-energy collisions with a coalescence picture, namely the formation of nuclei from nucleons emitted from the hot fireball which are close in phase space and "bind" to the later detected nucleus. Depending on the used phase space (sometimes only coordinate or momentum space, often also the combination of the two), the models can be rather successful in the description of the momentum spectra and the integrated production yields. The quality of the description also depends strongly on the nucleon spectra, normally the final state spectra, which are used in the coalescence approach. Two recent reviews highlight these approaches~\cite{Chen:2018tnh,Braun-Munzinger:2018hat}.
One usually argues, in particular in heavy-ion collisions, that the temperature where most of the particles are produced is much higher than the low binding energies of the nuclei, i.e. 2.2 MeV for the deuteron compared to the 156 MeV chemical freeze-out temperature $T_\mathrm{ch}$~\footnote{The chemical freeze-out temperature is typically defined as the temperature where the production yields of particles are fixed and only elastic collisions are still allowed, which then stop to happen at the so-called kinetic freeze-out temperature, where also (transverse momentum) spectra of particles are frozen.}. Nevertheless, the analysis of the production yields of all particles, including light nuclei, gives a similar temperature, and if only the light particles are used the prediction for the light nuclei is in good agreement with the measured production yields. In fact, if only the nuclei are used to extract a temperature the result is very close to the 156 MeV, namely $T_\mathrm{ch} \approx 160$\,MeV~\cite{Acharya:2017bso,Andronic:2017pug,Braun-Munzinger:2018hat}.

The description of the production of light nuclei using a statistical approach dates back to the first data from the CERN Proton Synchrotron where R. Hagedorn tried to describe the data from p-nucleus collisions~\cite{hagedorn60}. It is worth to notice, the "birth" of the coalescence model mentioned above is also connected with this data~\cite{butler_pearson61,butler_pearson63}. 
Nevertheless, the aforementioned model of Hagedorn is not a Hadron Resonance Gas (HRG) model as he introduced later on~\cite{Hagedorn:1965st,Hagedorn:1968zz,Hagedorn:1984uy}, but more a pure phase-space model.

From the partition function of the hadron resonance gas all thermodynamical quantities for hadrons and light (anti-)(hyper-)nuclei can be computed. Specifically, one can compute, for each hadron, its density $n(T,\mu,V)$. If all hadrons are produced from a state of thermodynamical equilibrium then, for a given data set, e.g. one beam or center-of-mass energy, the measured hadron yield for hadron $j$, ${\rm d}N_j/dy$ at a given rapidity $y$ but integrated over transverse momentum, should be
reproduced as ${\rm d}N_j/dy = V \cdot n(T,\mu,V)$. In practice, a fit is performed for each data set to the measured yields to determine the 3 parameters $T, \mu_B, V$. The potentials $\mu_Q$ and $\mu_S$ are fixed by strangeness and charge conservation.

Since the beginning of the 90s a very large body of data on hadron yields produced in ultra-relativistic nuclear collisions has been collected. From an analysis of these data in the spirit of the above approach convincing evidence has been obtained \cite{BraunMunzinger:2001ip,BraunMunzinger:2003zd,Becattini:2005xt,Andronic:2005yp,Stachel:2013zma,Becattini:2016xct,Andronic:2017pug} that the yields of all hadrons produced in central (nearly head-on) collisions can indeed be very well described, yielding the complete energy dependence of the parameters $T, \mu_B, V$ \cite{Becattini:2005xt,Andronic:2005yp}, see in particular also the recent fit to the precision LHC data \cite{Andronic:2017pug}. For 
recent reviews see \cite{Braun-Munzinger:2015hba,Andronic:2017pug}. Since the yields of particles are frozen at these parameters the corresponding temperature is also called chemical freeze-out temperature $T_{ch}$, as already indicated above.

The description of light nuclei in a HRG assumes that the nuclei are handled as normal hadrons, i.e. mesons or baryons, characterised mainly by their quantum numbers (spin $S$, isospin $I$, total angular momentum $J$, baryon number $B$, strangeness $S$, and charge $Q$) and mass. As a matter of fact, the very important feed-down in the baryon and meson sector seems to be negligible for the nuclei (at top RHIC and LHC energies). The population of the light nuclei states is mainly driven by their mass (or mass number A). The weakly decaying hypernuclei leading to additional yields for the light nuclei (of the same mass number A) are suppressed in the analysis since they are usually removed by a selection on topological quantities in the analysis. In addition, these decays have branching ratios and reconstruction efficiencies on the order of 10\% each, leading to a suppression of about 100 compared to the pure nuclei yields. The same holds true for hypernuclei decays into daughters of a mass number A-1 or lower.

\begin{figure}[!h]
\begin{center}
\includegraphics[width=0.8\textwidth]{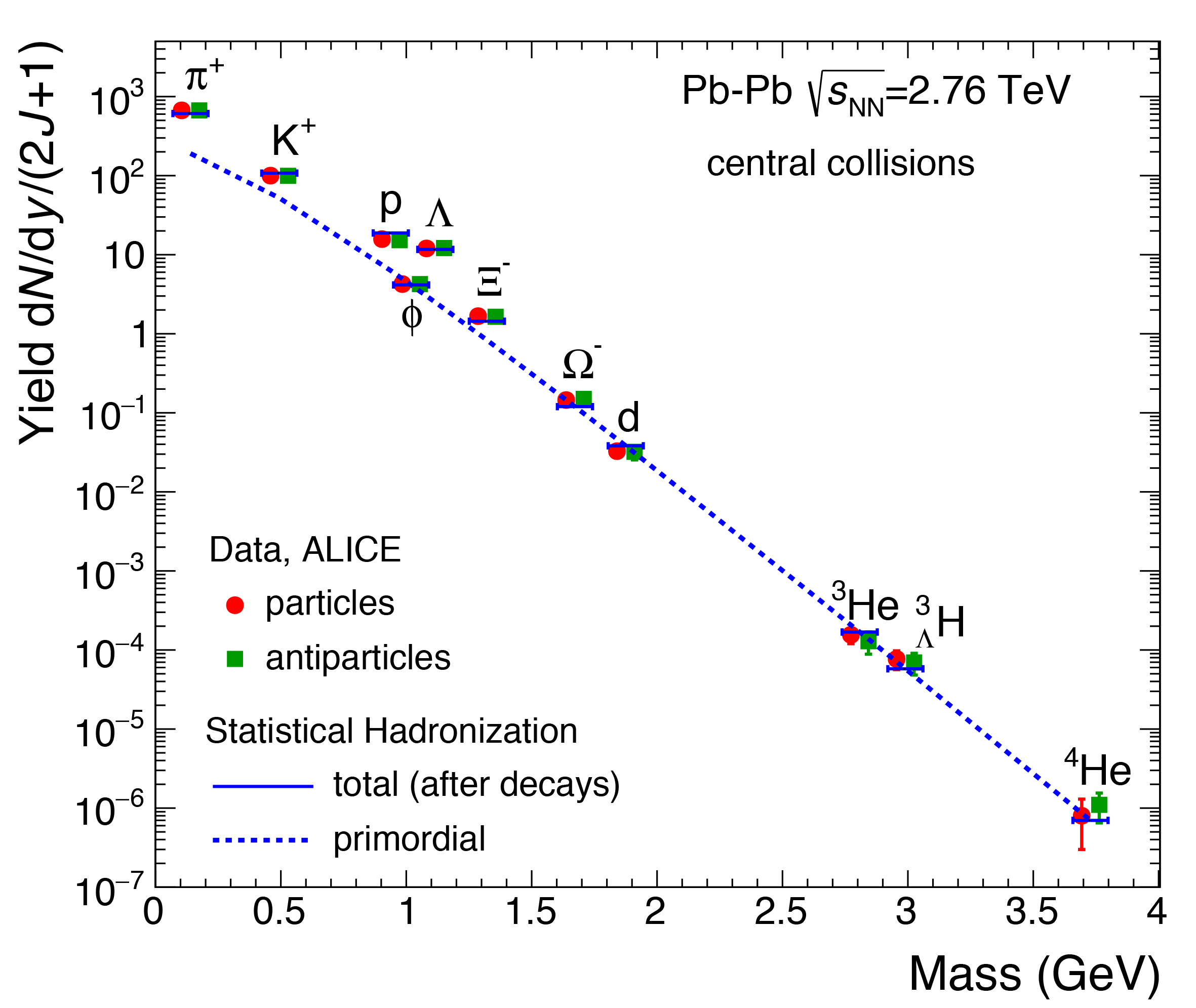}
\caption{\label{nature_plot} Thermal model description of the production yields (rapidity density) divided by (2$J$+1) as a function of mass for different particle species (antiparticles in green and particles in red) in heavy-ion collisions at the LHC (taken from~\cite{Andronic:2017pug}, where also more details can be found). The full lines correspond to the yields from the thermal model after decays happened, the dashed line shows the initial (primordial) yield.}
\end{center}
\end{figure}

These facts lead to a strong suppression (factor 330 at the LHC) of feed-down from higher mass states for instance seen in the ALICE data~\cite{Andronic:2017pug,Braun-Munzinger:2018hat,Andronic:2018vqh}. The previous details can be seen from the rather prominent plot displayed in Fig~\ref{nature_plot}, which shows the particle yields divided by (2$J$+1) as a function of mass. The measured yields agree very well with the full lines, which are taking into account the feed-down from higher mass states, whereas the dashed line corresponds to the primordial yields of the particles, i.e. the yields populated initially depending on the chemical freeze-out temperature $T_{ch}$, the baryo-chemical potential $\mu_{B}$ and the volume $V$.  

Important work was done in the beginning to middle of the 80s to describe the data on light nuclei production from the Bevalac. Back then the whole entropy production was connected with baryons and nuclei, since only a small amount of pions is created in comparison with the high-energies. In fact, the entropy was found to be directly extractable from the $d/p$ ratio and connected through the formula $S = 3.95 - d/p$~\cite{Siemens:1979dz}, whereas the data back then gave values of about 5 to 6. This was understood by hydrodynamic calculations which included decays from particle unstable excited nuclei~\cite{Stoecker:1984py}.
Light nuclei yields at these energies have strong contributions from the decay of intermediate mass fragments which are produced in excited states or even only exist as a typically broad resonance state as pointed out and used by Hahn \& St\"{o}cker~\cite{Hahn:1986mb,Hahn:1986pw}. This work has been re-discovered recently and some work is done to describe the data as SIS18 energies taken with the HADES experiment and the data from STAR from the beam-energy scan at RHIC~\cite{Vovchenko2019}. These nuclei resonances are also important in a recent pre-clustering approach~\cite{Shuryak:2019ikv}.

\begin{figure}[!h]
\begin{center}
\includegraphics[width=0.65\textwidth]{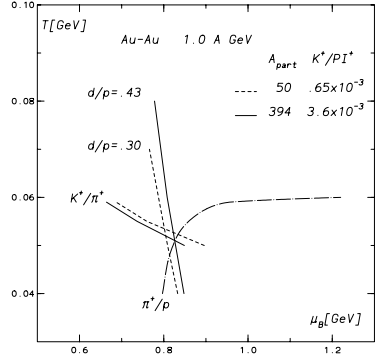}
\caption{\label{cleymans_T_mu} Freeze-out parameters $T$ vs. $\mu_B$ for different particle ratios and number of participants in the collision $A_\mathrm{part}$ (from~\cite{Cleymans:1998yb}). }
\end{center}
\end{figure}

It was observed rather early that including the nuclei in the fit can constrain the fits much better and at the AGS and SPS not all particle species have been measured by single experiments, thus it was important that another experiment measures other particles to assess a thermal model fit.  
Figure~\ref{cleymans_T_mu} for instance shows the lines in the $(T-\mu_B)$ plane corresponding
to the measured particle ratios in Au-Au collisions at 1 $A\cdot$GeV at SIS18 energies at GSI.
All lines have a common crossing point  around $T\sim 50$ and
$\mu_B\sim 822$ MeV. A value for the radius $R\approx 6.2$ fm (corresponding to a volume $V$ of about 1000~fm$^{3}$)
is needed to describe the measured $K^+/\pi^+$ ratio with the
freeze-out parameters
 extracted for $\pi^+/p$, $\pi^+/\pi^-$ and $d/p$.
If the d/p ratio is not used in the extraction of these parameters the favoured $(T-\mu_B)$ value can be very different. A similar observation was made by the HADES Collaboration, while describing their Au--Au data~\cite{HADES2019}.

\section{Recent developments}

\subsection{Statistical hadronisation model}

\begin{figure}[!htb]
\begin{center}
\includegraphics[width=0.9\textwidth]{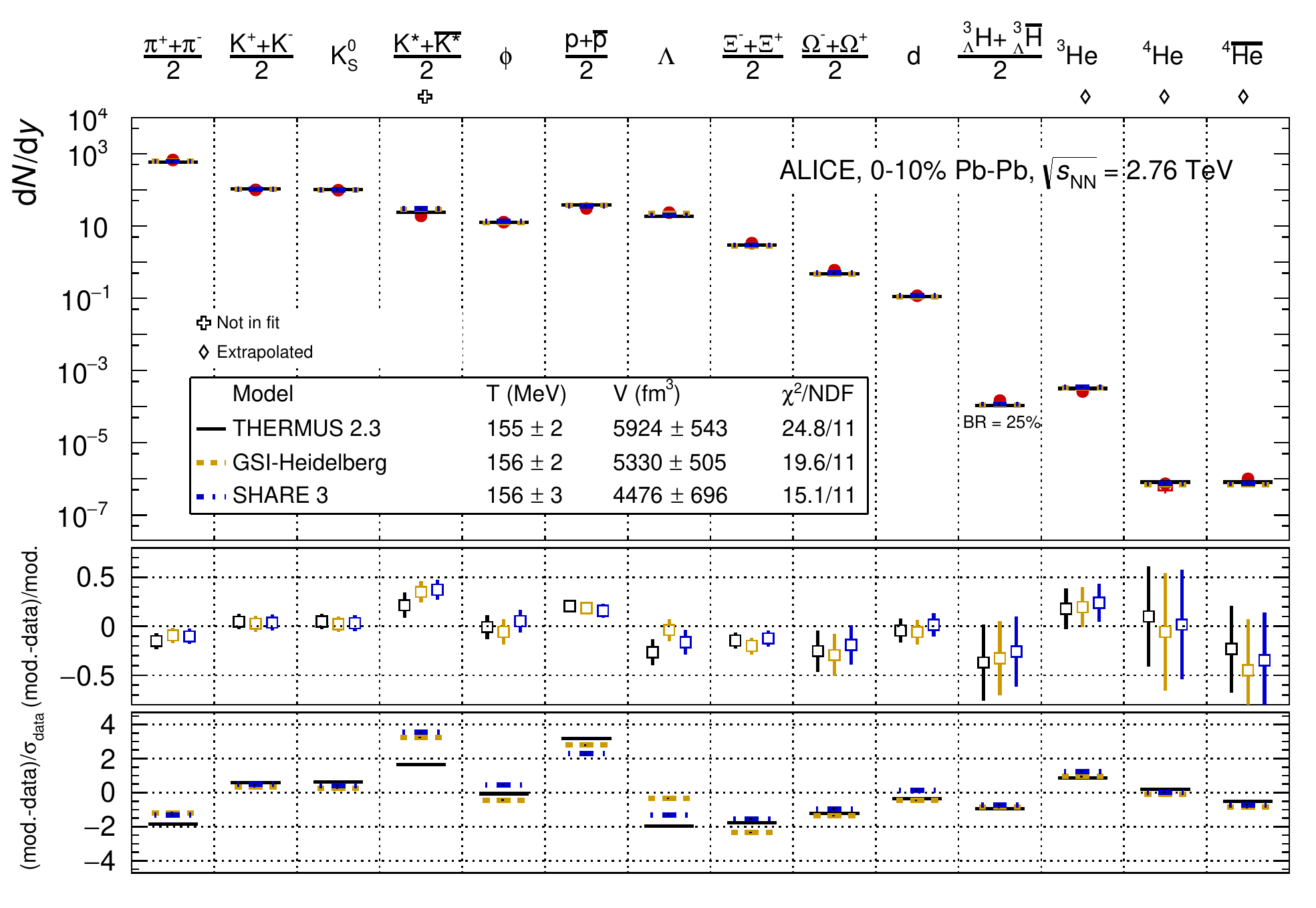}
\caption{\label{thermal_fit_lhc} Thermal model description of the production yields (rapidity density) of different particle species in heavy-ion collisions at the LHC for a chemical freeze-out temperature of about 156 MeV using three different thermal model implementations (from~\cite{Acharya:2017bso}).}
\end{center}
\end{figure}

In the following we compare the hadron production at the LHC with the thermal model. In Figure~\ref{thermal_fit_lhc} the result is shown for a thermal model analysis of the data collected by the ALICE Collaboration using the GSI-Heidelberg model~\cite{thermalModel,pbm,pbm1,anton_thermal,Stachel:2013zma,anton_sqm2016,Andronic:2017pug}, the THERMUS package~\cite{Wheaton:2004qb,Wheaton:2004vg} and SHARE~\cite{rafelski0,rafelski1,Torrieri:2004zz}. Very good agreement is obtained for $T_{chem} = 156$ MeV over the nine orders of magnitude in particle production yields. 

It is worth to mention, that THERMUS and SHARE are publicly available codes and allow for a non-equilibrium treatment in the fit. They both allow for strangeness under-/over-saturation with the $\gamma_\mathrm{s}$ factor, whereas SHARE allows in addition also the light quark under-/over-saturation with the $\gamma_\mathrm{q}$ factor. In THERMUS a canonical treatment is possible by introducing a correlation radius $R_C$, in which the conservation of the corresponding quantum number is conserved explicitly. The most recent developed thermal model package FIST~\cite{ThermalFIST,Vovchenko:2019pjl} offers the aformentioned features and some additional ones. It is rather versatile and can be used in different applications, e.g. as a tool for future analyses that would include canonical suppression, eigenvolumes, fragment feeddown, a hadronic phase or Saha equation effects and multiple freeze-out scenarios, basically all the different aspects discussed in this review.

In the context of non-equilibrium models, something interesting was observed when nuclei were first included in the thermal model fits of the ALICE data. If one leaves $\gamma_\mathrm{s}$ and $\gamma_\mathrm{q}$ free and fits only hadrons and no nuclei, one gets $T_{chem} = 138$ MeV,  $V = 3100$ fm$^{3}$, $\gamma_\mathrm{s} = 2.01$ and $\gamma_\mathrm{q} =1.63$; whereas if the nuclei are included the fit gets very close to the equilibrium fit (fixing $\gamma_\mathrm{s}$ and $\gamma_\mathrm{q}$ to unity)~\cite{Floris2014103}.

As stated above, at LHC energy, the baryo-chemical potential $\mu_B$ which is a measure of the difference of production probabilities for baryons and anti-baryons is expected to be close to zero, since the LHC centre-of-mass energy exceeds twice the baryon mass by more than a factor of $10^3$. The value for the fit presented in the figure from the fit  is in excellent agreement with this expectation. The nearly vanishing baryo-chemical potential leads to equal production yields of baryons and anti-baryons and in consequence also to equal yields of nuclei and anti-nuclei for the different species.  

This also implies that measurements of particle production at LHC energies are relevant for the understanding of the evolution of the early universe. In fact, different from the situation for nuclear collisions at LHC energy, the production of nuclei in the early universe can not occur when the baryons are produced because the photons, which are still in equilibrium with the baryons, would destroy all formed nuclei immediately. Thus, the formation of nuclei occurs in the early universe at a much later time after the temperature has dropped sufficiently, such that no thermal photons are left to destroy the formed deuterons. From this point on, the process n + p $\rightarrow$ d + $\gamma$ is dominating the detailed balance of the two processes, deuterons are produced and the backward reaction is energetically suppressed.

\begin{figure}[!h]
\begin{center}
\includegraphics[width=0.49\textwidth]{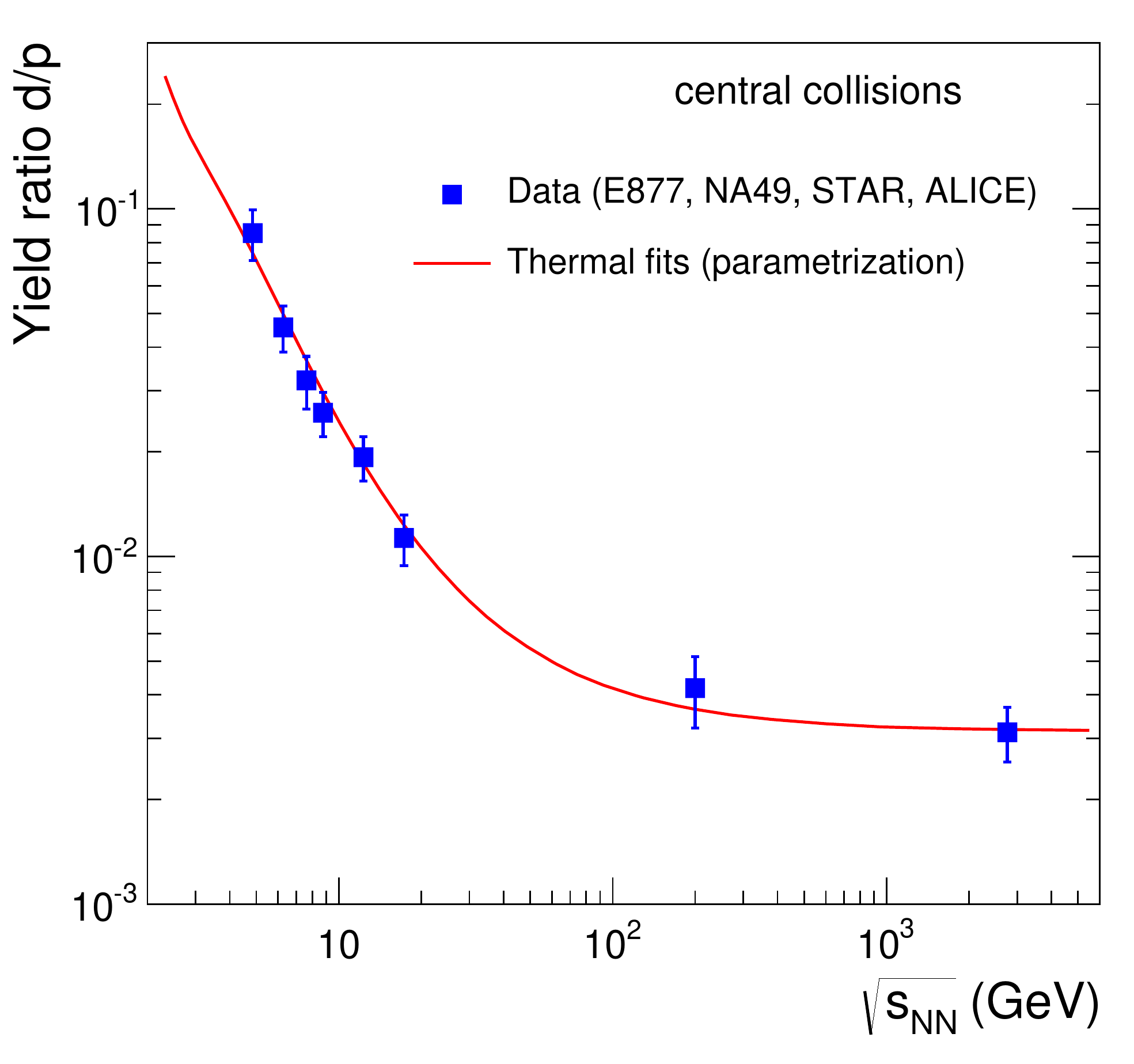}
\includegraphics[width=0.49\textwidth]{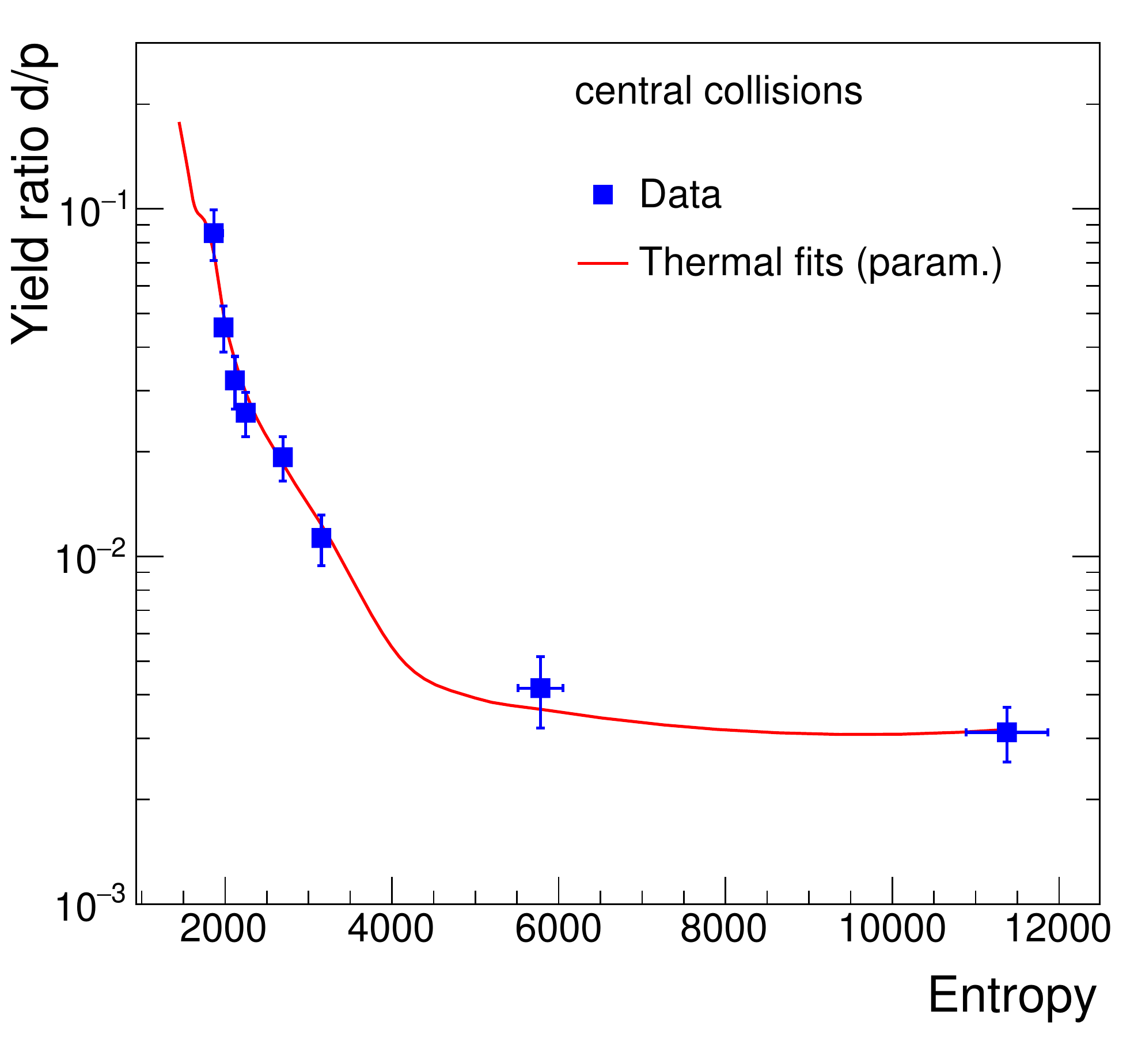}
\caption{\label{d_to_p_sqrts} Deuteron-to-proton ratio as measured in central nuclear collisions at different centre-of-mass energies $\sqrt{s_{\mathrm{NN}}}$ (left panel). Deuteron-to-proton ratio as measured in central nuclear collisions versus produced entropy (right panel). The data points are compared with predictions of the ratio and entropy based on the thermal model (parameterised in the red line). As shown in~\cite{Braun-Munzinger:2018hat}.}
\end{center}
\end{figure}

Since, in this review, we are in particular interested in loosely-bound states we show in Figure~\ref{d_to_p_sqrts} the deuteron-to-proton ratio in relativistic nuclear collisions as a function of centre-of-mass energy, bridging data from the SPS to RHIC to the LHC. Assuming thermal production of deuterons according to its mass and spin reproduces the data very well, implying that the statistical hadronisation model is a useful tool to estimate production yields also for loosely-bound states as developed in \cite{pbm,pbm1,thermalModel}.
The application of the parameterization of the energy-dependence of $T_{chem}$ and $\mu_B$~\cite{Becattini:2005xt,Andronic:2005yp} within the framework of the statistical hadronisation model leads to an impressive description of all hadron production data. In fact, yields for the production of loosely-bound states at LHC energy were successfully predicted in  \cite{thermalModel} based on the statistical hadronisation model before data taking. This shows that the production of nuclei is quantitatively well reproduced within the framework of the statistical hadronisation model, implying that the same parameters ($T_{chem}, \mu_B, V$) governing light hadron production yields also determine the production of light composite objects, with only their mass and quantum numbers and not structural parameters such as binding energy or radius as input.
 
Another way to look at the deuteron-proton ratio is displayed in Figure~\ref{d_to_p_sqrts} extracted from the thermal model~\cite{Andronic:2005yp}. In this Figure, the d/p ratio is shown as function of the entropy per unit of rapidity in the collision. As naively expected, increasing the entropy leads first to a precipitous drop of the ratio, as the entropy/baryon scales $\propto -\ln{(\mathrm{d/p})}$ \cite{Siemens:1979dz,Nagamiya:1984vk}. Above $\sqrt{s_{\mathrm{NN}}} \approx 20$ GeV the chemical freeze-out temperature saturates at around 160 MeV, implying that the entropy density stays constant. The main entropy increase is then due to the volume expansion of the fireball at freeze-out, implying that the d/p ratio approaches a constant value of $\approx  3 \cdot 10^{-3}$.
 
 \subsection{Multiple chemical freeze-out scenarios}

The first results from multi-strange particle production at RHIC indicated some tension for thermal model fits and also when the transverse momentum spectra of (multi-)strange particles are compared with those of other light-flavoured hadrons that can not be easily described by one set of freeze-out parameters (see for instance~\cite{Fachini:2006ch}).
One possible idea to overcome this tension would be a multiple freeze-out scenario, in the easiest case with just two different temperatures, allowing $u+d$ quarks to freeze-out later than $s$ quarks, namely strangeness is freezing out at a higher temperature than the lighter quarks. This can also be supported by lattice QCD studies and results on net-kaon fluctuations from the beam-energy scan at RHIC~\cite{Bellwied:2013cta,Alba:2015iva,Bellwied:2017uat,Ratti:2019ytu}. The net-kaon fluctuations at the highest energies would favour a freeze-out temperature of about 10-15 MeV above the one extracted from the usual thermal model fits~\cite{Ratti:2019ytu}. 
 
 \begin{figure}[!htb]
 \begin{center}
 \includegraphics[width=0.85\textwidth]{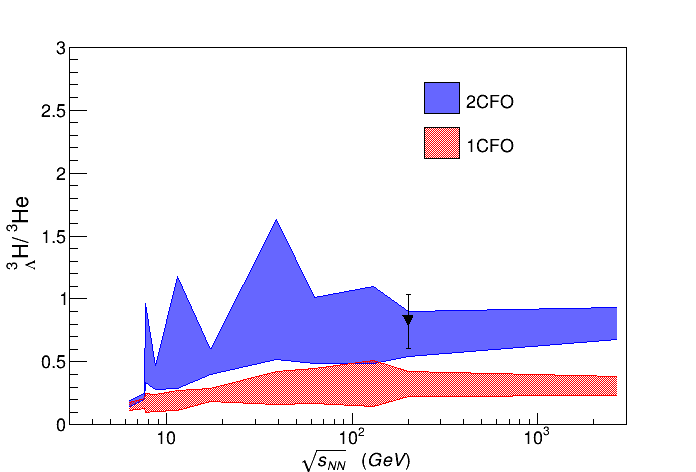}
 \caption{\label{hypt_helium} $^{3}_{\Lambda}$H/$^{3}$He ratio as function of the centre-of-mass energy per nucleon pair using a single freeze-out scenario, named 1CFO (red band), and for the two freeze-out scenario, indicated as 2CFO (blue band). The bands reflect the uncertainties associated with the ratios extracted in the thermal model fits. In addition, the measurement of STAR\cite{star} is shown as black marker. Figure from~\cite{Chatterjee:2014ysa}.
 For comparison, the published $^{3}_{\Lambda}$H/$^{3}$He ratio measured by the ALICE Collaboration at 2.76 TeV is $ 0.47 \pm 0.10 (\mathrm{stat.}) \pm 0.13 (\mathrm{stat.}$)~\cite{hypertriton}.
 }
 \end{center}
 \end{figure}
 
Similar conclusions are reached in studies allowing for two decoupled freeze-outs involving light (anti-)(hyper-)nuclei like presented in~\cite{Chatterjee:2013yga,Chatterjee:2014ysa,Chatterjee:2015fua,Chatterjee:2016dld}. This approach solves the result presented in the paper on the discovery of the anti-hypertriton~\cite{star}, where already the hypertriton to $^{3}$He ratio ($^{3}_{\Lambda}$H/$^{3}$He) and the corresponding ratio of the anti-particles is shown. These experimental results are still not possible to describe easily within a single freeze-out scenario, whereas the two freeze-out scenario is able to describe these data, as shown in Fig.~\ref{hypt_helium}. 
The ratio measured by STAR is clearly in contradiction with the results from ALICE at higher energies, where the yields (and the corresponding ratio) is well described by the statistical-thermal model. Thus it would be beneficial if the yields could be re-measured by the STAR Collaboration, since in principle an energy dependence is not expected.

\begin{figure}[!htb]
\begin{center}
\includegraphics[width=0.7\textwidth]{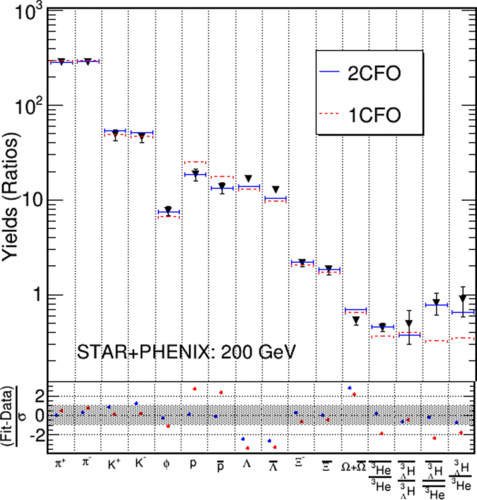}
\caption{\label{star_phenix} Particle yields and ratios (sometimes the experiments only provide ratios and not the yields) from the STAR and PHENIX collaborations at $\sqrt{s_\mathrm{NN}} = 200$~GeV compared with the single freeze-out scenario, named 1CFO and presented as the red dashed lines, and for the two freeze-out scenario indicated as 2CFO (blue line). The lower panel shows the difference between thermal model fit value and the experimental values compared to the experimental uncertainty (indicated as $\sigma$). Figure from~\cite{Chatterjee:2014ysa}.}
\end{center}
\end{figure}

A full fit using the one and two freeze-out scenarios for the available data from the STAR and PHENIX collaborations at $\sqrt{s_\mathrm{NN}} = 200$~GeV is displayed in Fig.~\ref{star_phenix}. As from the net-kaon fluctuations mentioned above also here the difference of the two temperatures extracted in the two freeze-out scenario is about 10 MeV (155 compared 163 or 162, depending if nuclei are included in the fit or not). It is worth to mention, that in~\cite{Chatterjee:2014ysa} also an estimate comparing with the coalescence model coupled to the two freeze-out scenario (forming (anti-)(hyper-)nuclei from (anti-)baryons formed from at the chemical freeze-outs) is done and the authors conclude that also in this scenario it is not easy to distinguish between the coalescence and thermal model.

A summary of these findings also for lower and higher energies completing the picture by comparing in addition data from AGS, SPS and LHC can be found in a comprehensive review~\cite{Chatterjee:2015fua} on this topic, focusing not only on light nuclei.  

It is worth to mention, that the picture of different freeze-outs and thus different phase boundaries for different quark species calls the concept of the QGP in question, namely how should hadrons 'hadronise' when they contain a mixture of $u, d, s$ quarks. The discussed fits show even less tension for the $sss$-state, i.e. the $\Omega^{-}$ hyperon, which contradicts the sequential freeze-out slightly.
In addition, the charm quark seems to be well described by the same temperature as the light quarks, as shown in~\cite{Andronic:2019wva}, where the only additional input to the statistical hadronisation model is the charm cross-section, or better the number of $c\bar{c}$-pairs.

\subsection{Influence of the eigenvolume}
Generally, in ideal gas statistical-thermal model approaches the particles are taken to be point like. In a statistical mechanics course the repulsive interaction is usually first introduced via van der Waals forces. This is also an approach introduced in the description of the extended volume in the bootstrap model calculations presented in~\cite{Hagedorn:1980kb}. For the thermal model, this implies that one applies an excluded volume model as discussed in~\cite{Rischke:1991ke}, replacing the volume $V$ by $V - bN$ and by that modifies the pressure of the ideal gas. Significant work has been done by Vovchenko et al.~\cite{Vovchenko:2015cbk,Vovchenko:2016ebv,Vovchenko:2016eby,Vovchenko:2016mwg,PhysRevC.95.024902,Vovchenko:2017zpj,Vovchenko:2016mwg,Vovchenko:2016rkn,Alba:2016hwx,Vovchenko:2017drx,Vovchenko:2017ygz,Poberezhnyuk:2017yhx}.

The modification of the volume can be easily seen from Fig.~\ref{thermal_fit_lhc}, where two models have values of the Volume above 5000 fm$^{3}$ and one gives a volume of only 4500 fm$^{3}$. GSI-Heidelberg and THERMUS are including an eigenvolume of the particles (baryon radius r$_\mathrm{B}$ and meson radius r$_\mathrm{M}$) of 0.3 fm), whereas for SHARE the particles are treated to be point-like.

\begin{figure}[!htb]
\begin{center}
\includegraphics[width=0.7\textwidth]{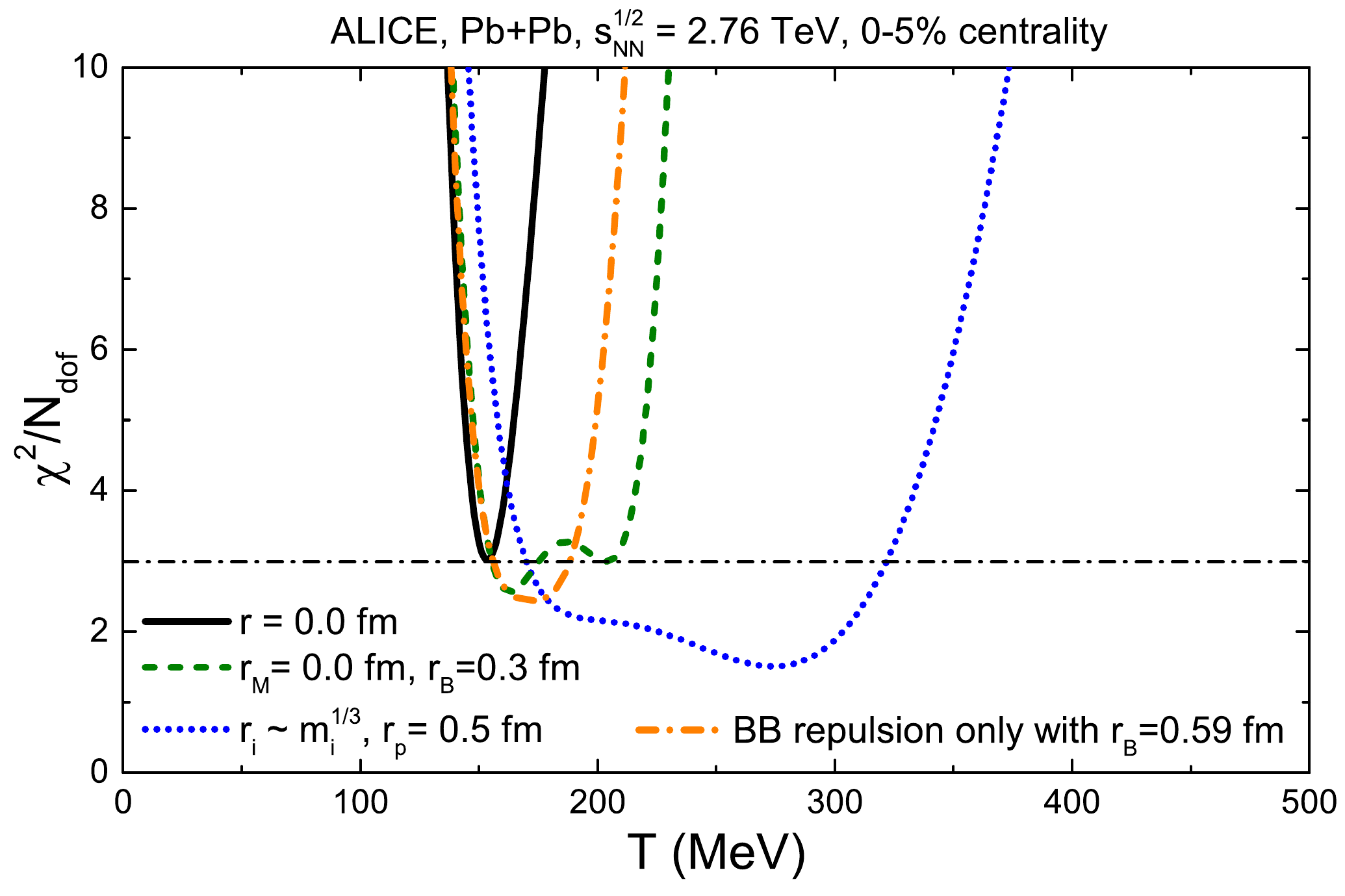}
\caption{\label{alice_chi2} Distribution of the $\chi^{2}$ per degree of freedom (N$_\mathrm{dof}$) as function of the temperature using different assumptions for the radius of hadrons: point-like particles (black), baryon radius r$_\mathrm{B}$ of 0.3 fm and meson radius of zero (dashed green line), scaling of radius with mass for all particles and a fixed proton radius of 0.5 fm (blue dashed line), and a particular baryon-baryon interaction with a radius of 0.59 fm. Figure from~\cite{Vovchenko:2015cbk}. }
\end{center}
\end{figure}

In Fig.~\ref{alice_chi2} the $\chi^{2}$ per degree of freedom is shown for different assumptions in thermal model fits of central ALICE data (as shown in Fig.~\ref{thermal_fit_lhc}). The change of the size of the fitted objects leads to a shift of the minimum and allows for a second minimum, where hadrons are supposed to be deconfined (according to lattice QCD results - no hadrons above $T \approx 170$ MeV).

\begin{figure}[!htb]
\begin{center}
\includegraphics[width=0.7\textwidth]{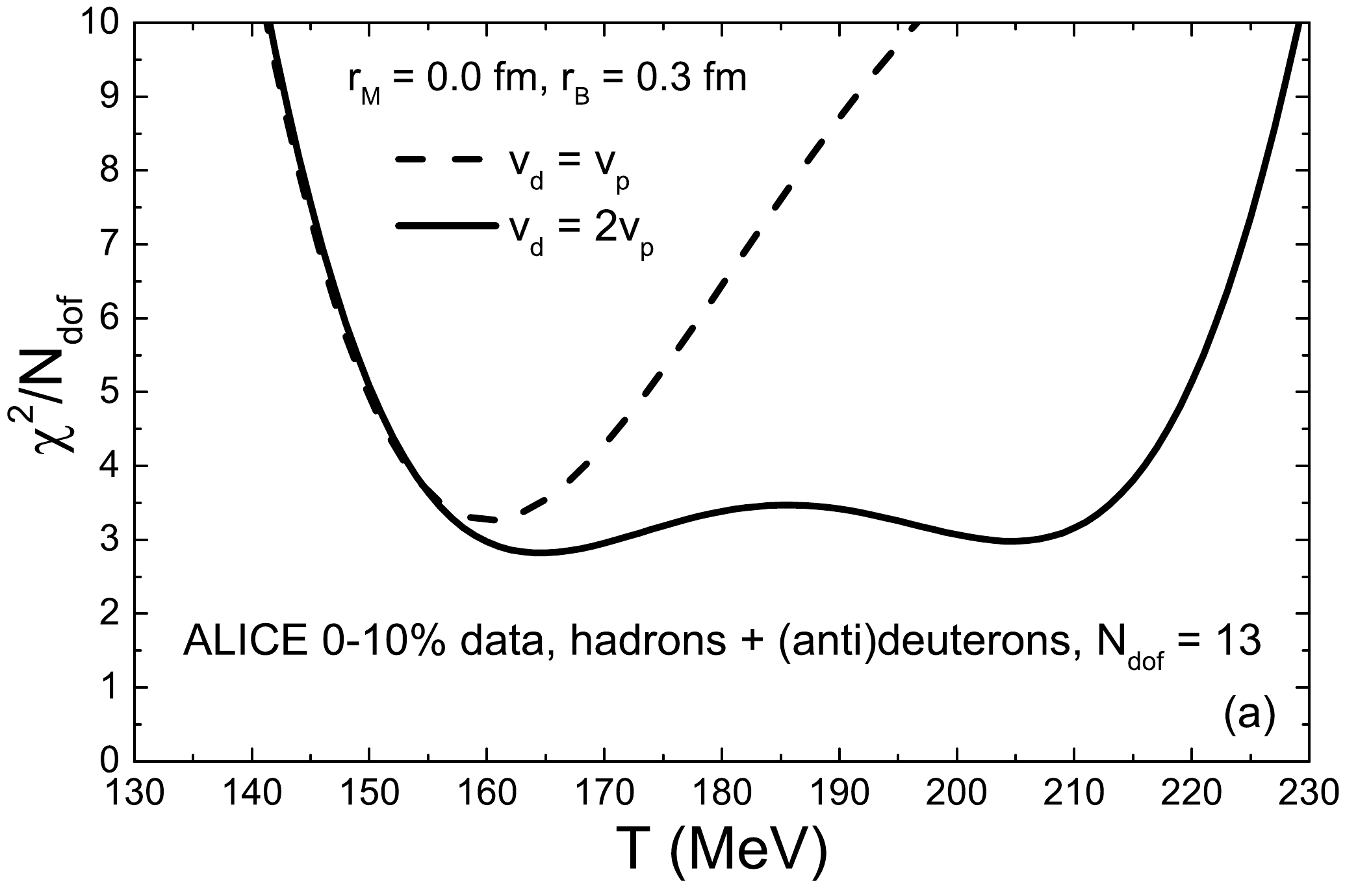}
\caption{\label{alice_chi2_d} Distribution of the $\chi^{2}$ per degree of freedom (N$_\mathrm{dof}$) as function of the temperature using different assumptions for the radius of hadrons: proton radius equal to the deuteron radius (full black line), and deuteron is twice as big as the proton. As shown in~\cite{Vovchenko:2016mwg}.}
\end{center}
\end{figure}

Figure~\ref{alice_chi2_d} shows the same observable as Fig.~\ref{alice_chi2}, but assumes a fixed scaling of the volume of nuclei by the baryon number (deuteron is as big as two protons). One observes that the minimum is shifted and a second minimum above 200 MeV is visible.

These studies show that thermal fits are very sensitive to the details of the modeling of the eigenvolume interactions and that if light
nuclei are included into thermal fits the results are even more sensitive to the assumptions regarding their eigenvolume parameters.

A different approach to taking into account the repulsion between hadrons is given by the S-Matrix formulation of statistical mechanics which is using the measured phase shifts of particles as input to include pion-nucleon interactions. This approach is able to resolve the aforementioned tension in the thermal model fit for the proton yields and leads to a better description of the data~\cite{Andronic:2018qqt}.

In fact, the inclusion of eigenvolume corrections for nuclei is still an open question in the treatment of hadron yields. 

\subsection{Canonical statistical approach}
In the standard grand-canonical description all investigated charges are conserved on average, but can fluctuate from one event to another.
This grand-canonical treatment of particle yields is appropriate when the number of produced particles and thus the volume (and/or temperature or baryo-chemical potential) is sufficiently large.
However, when the reaction volume is small, i.e. when the number of particles with carrying particular conserved charge(s) is of the order of unity or smaller, then the canonical treatment of the corresponding conserved charge(s) is necessary~\cite{Rafelski:1980gk,Hagedorn:1984uy,Cleymans:1990mn}.
In the canonical ensemble the conservation laws are exactly enforced from one event to another, which results in the so-called canonical suppression in the yields of particles carrying conserved charges relative to their grand-canonical values. The effect is stronger for the multi-charged particles, such as the multi-strange hyperons or light nuclei.
The canonical ensemble formulation of the thermal model has been successfully used to describe hadron abundances measured in small systems, including those created in such 'elementary' collisions as $e^+ e^-$~\cite{Becattini:1995if,Andronic:2008ev,Becattini:2008tx}, $pp$ or $p(\bar p)$~\cite{Becattini:1997rv,Becattini:2005xt,Becattini:2010sk}.

Canonical suppression effects have previously been considered at LHC energies for strangeness only~\cite{Kraus:2008fh,Adam:2015vsf,Vislavicius:2016rwi}.
A qualitative description of the multiplicity dependence of ratios of yields of various strange hadrons to pions was obtained in this strangeness-canonical ensemble picture~\cite{Adam:2015vsf,Vislavicius:2016rwi}. Whereas in~\cite{Vovchenko:2018fiy}, a full canonical treatment of baryon number, electric charge, and strangeness is applied and expected to influence the yields of light nuclei, especially the baryon number given that light nuclei carry multiple baryon charges.

\begin{figure}[!htb]
\begin{center}
  \includegraphics[width=.47\textwidth]{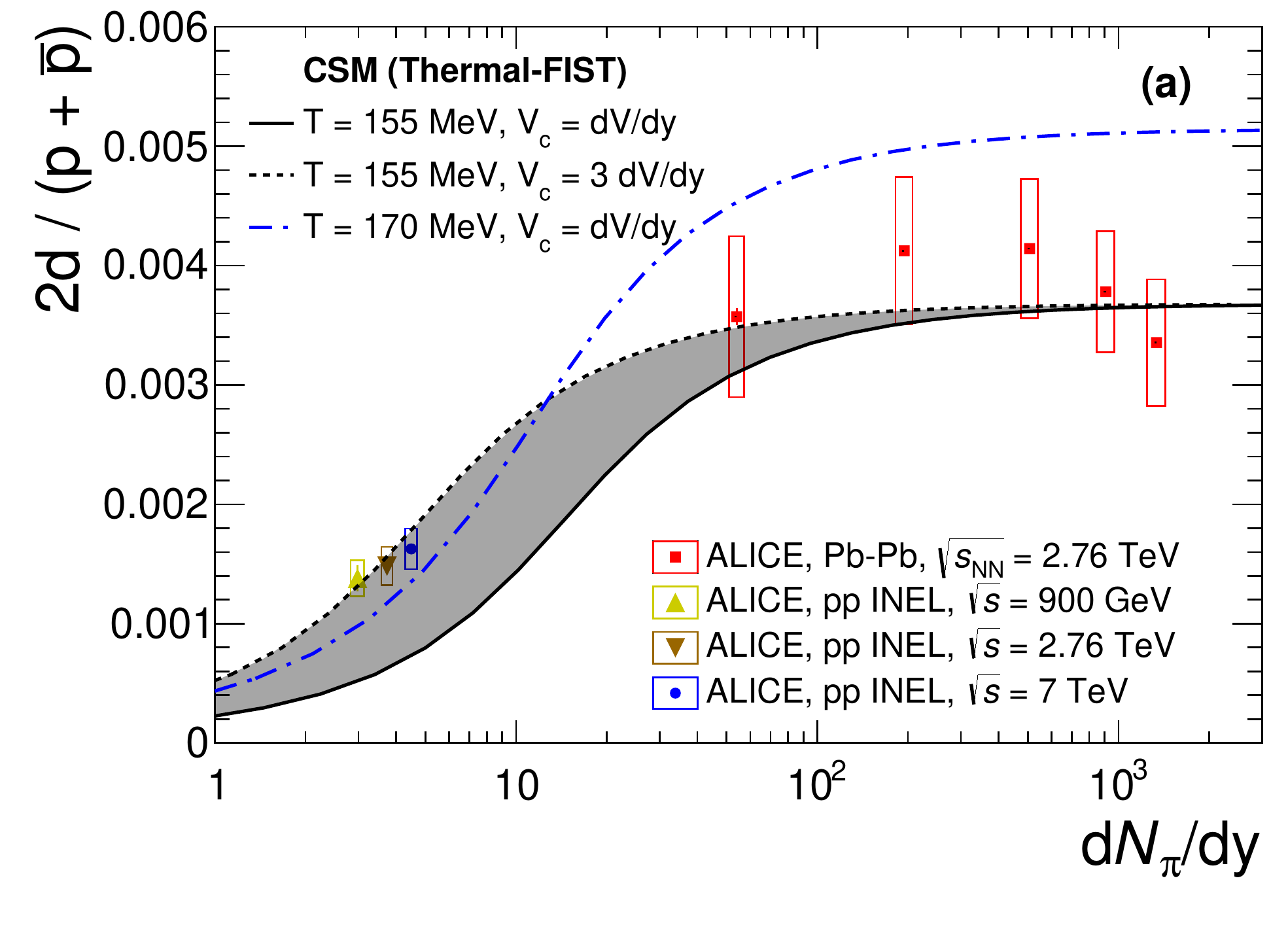}
  \includegraphics[width=.47\textwidth]{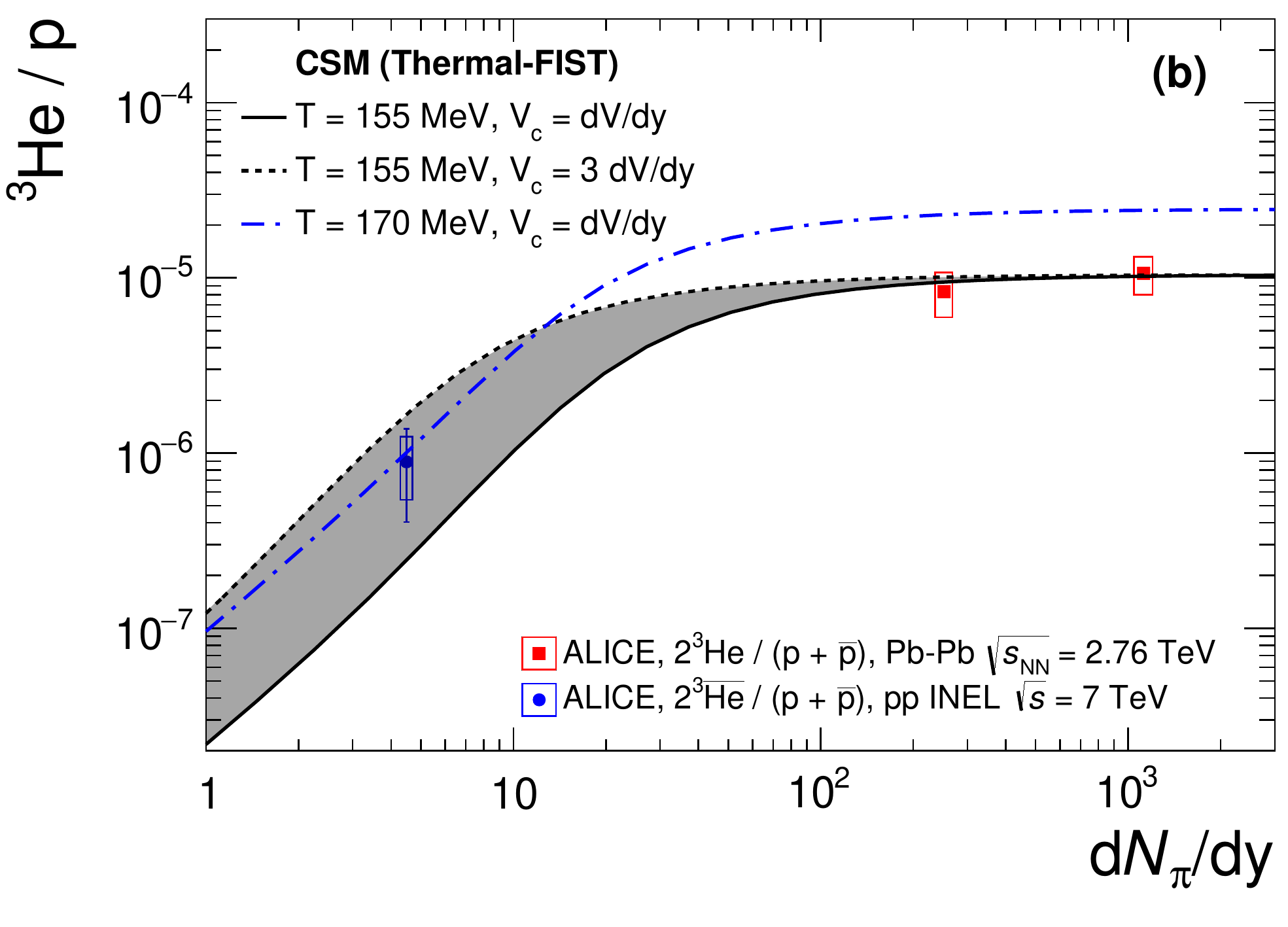}
  \includegraphics[width=.47\textwidth]{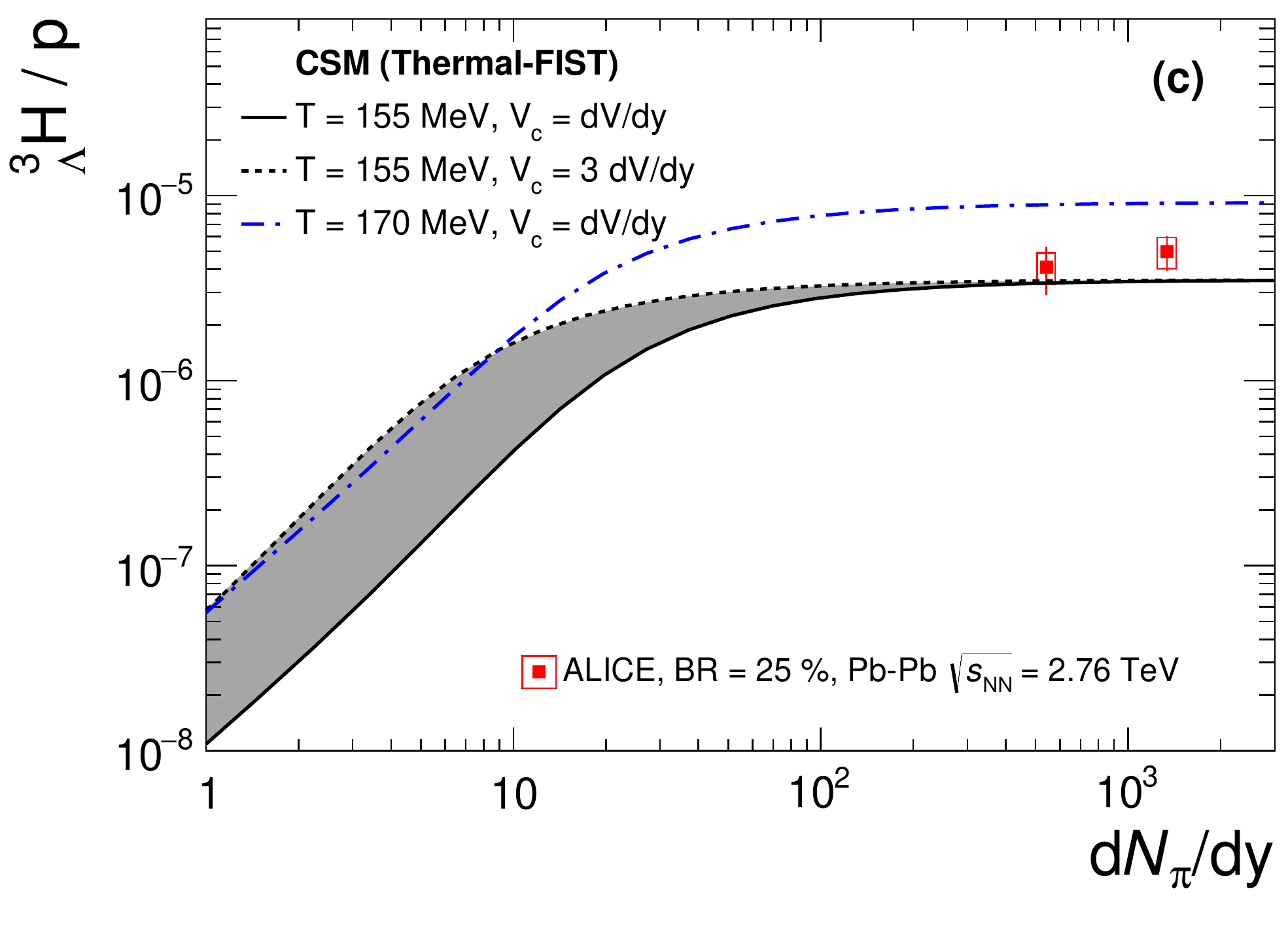}
  \includegraphics[width=.47\textwidth]{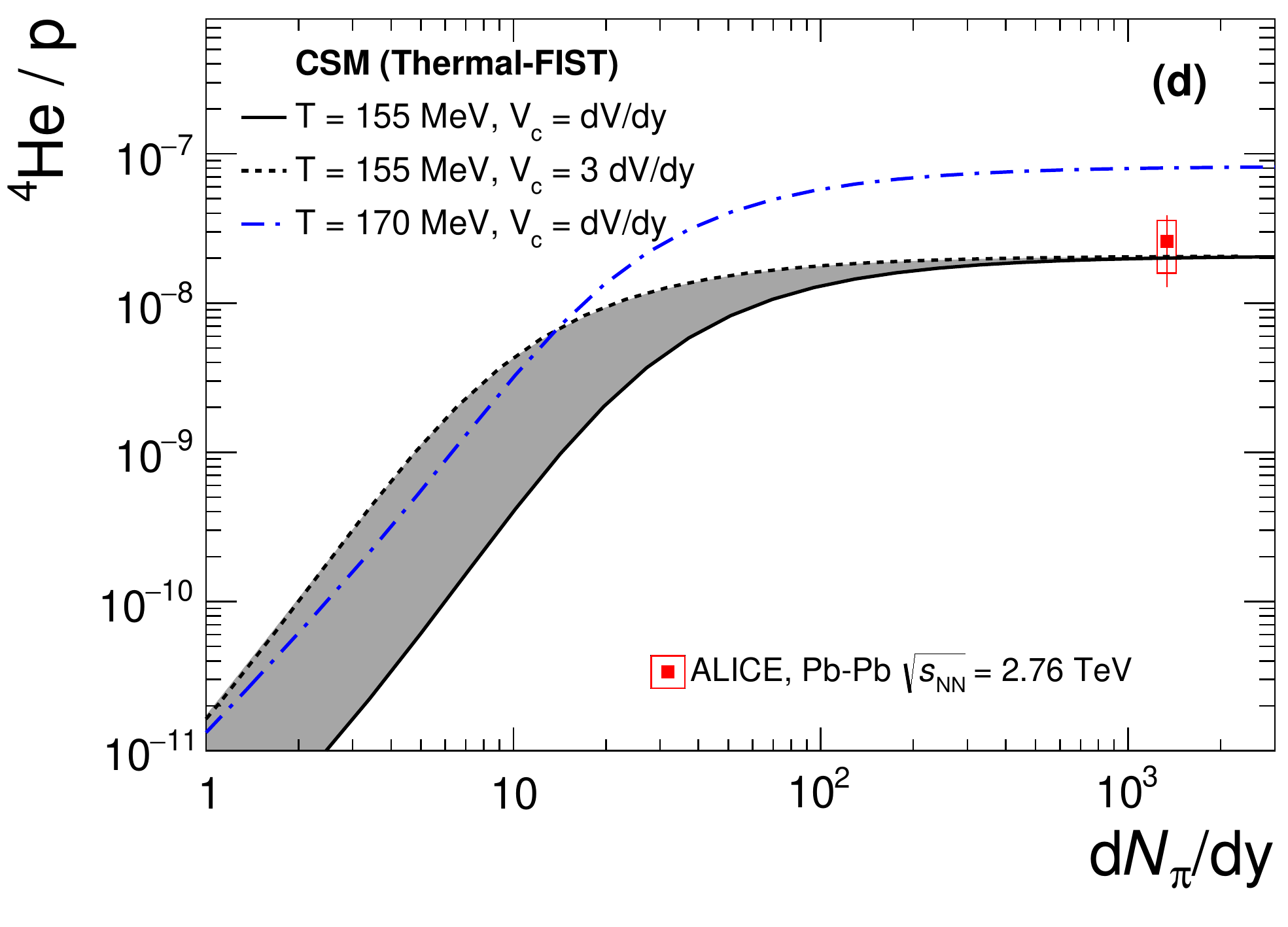}
  \caption{\label{d_over_p_dNdpi} Charged particle multiplicity dependence ($dN_\pi/dy$) of (a) $d/p$, (b) $^3\text{He} / p$, (c) $^3_{\Lambda} \text{H} / p$, and (d) $^4\text{He} / p$ ratios calculated in the canonical ensemble HRG model at $T = 155$~MeV for $V_c = dV / dy$~(solid lines) and $V_c = 3 \, dV / dy$~(dashed lines).
  Experimental data of the ALICE collaboration are shown where available~\cite{Adam:2015vda,Adam:2015yta,Acharya:2017bso,Acharya:2017fvb}. From~\cite{Vovchenko:2018fiy}.}
\end{center}
\end{figure}

A rather good description of the available data is reached if a chemical freeze-out temperature of 155 MeV is used as visible from the ratios depicted in Fig.~\ref{d_over_p_dNdpi}. As discussed in~\cite{Vovchenko:2018fiy}, the volume in which the charges are conserved exactly is unfortunately a bit arbitrary~\cite{Castorina:2013mba}. Therefore, the predicted trend as a function of multiplicity is given for two values of the correlation volume ($V_c = dV/dy$ and $V_c = 3 dV/dy$).
 
\subsection{Connection to the early universe}
\begin{figure}[!htb]
\begin{center}
\includegraphics[width=0.47\textwidth]{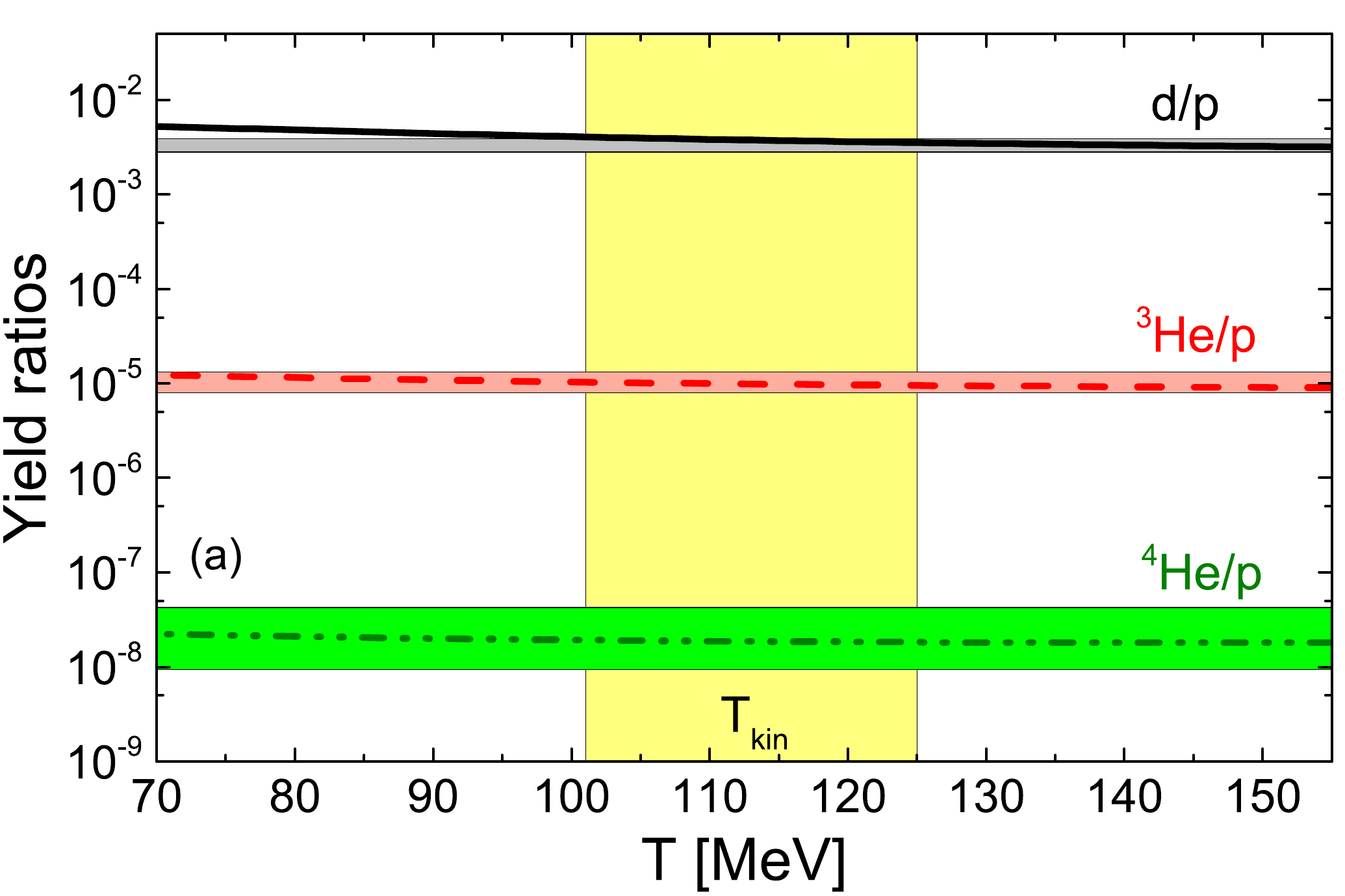}
\includegraphics[width=0.47\textwidth]{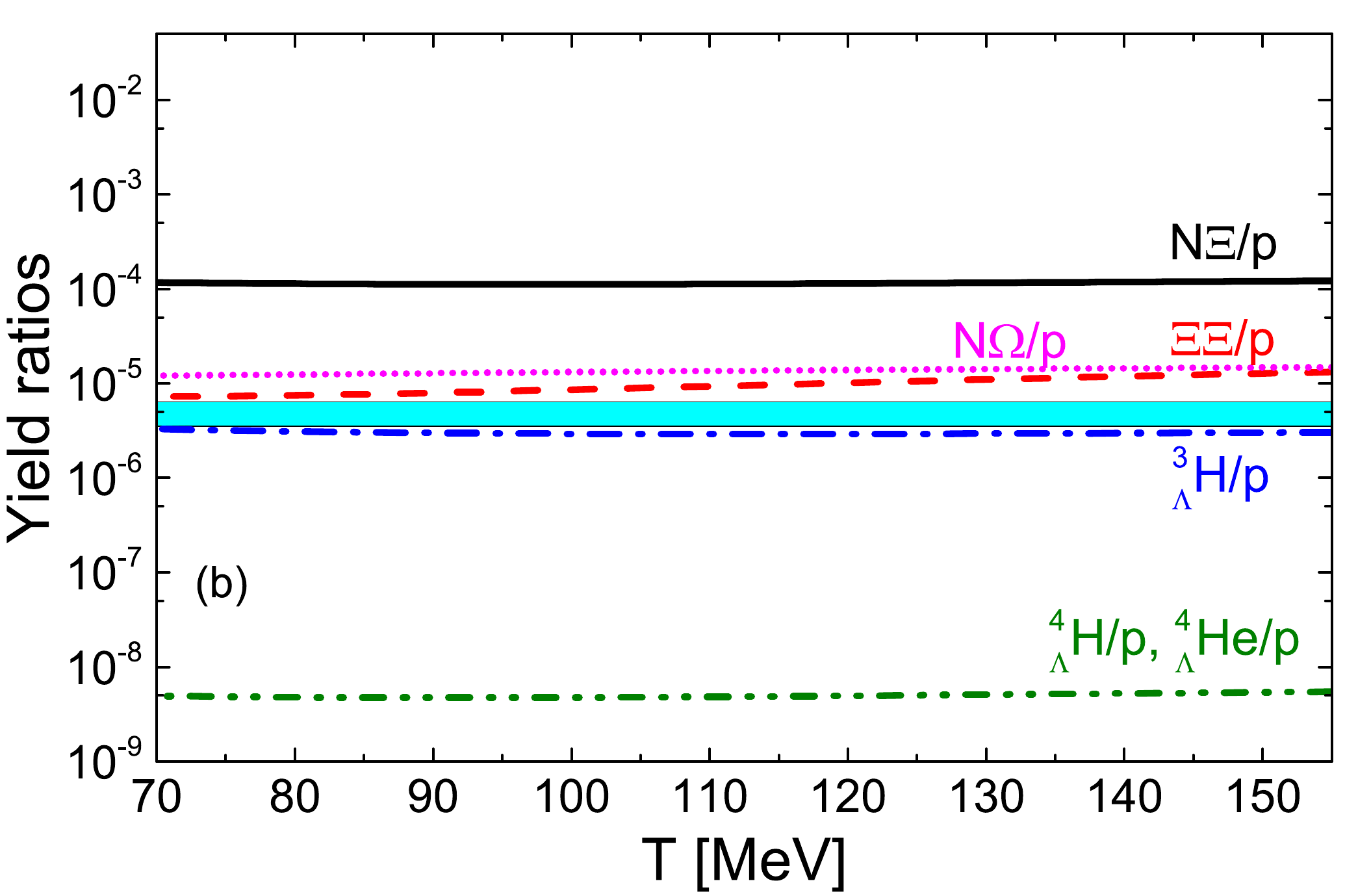}
\caption{\label{saha_2}Temperature dependence of yields ratios (a) d/p~(solid black line), $^3$He/p~(dashed red line), $^4$He/p~(double-dot-dashed green line), and (b) $N\Xi$/p~(solid black line),  $N\Omega$/p~(dotted magenta line),
   $\Xi\Xi$/p~(dashed red line),
   $^3_\Lambda$H/p~(dot-dashed blue line), and $^4_\Lambda$H/p and $^4_\Lambda$He/p~(double-dot-dashed green line), evaluated at $T<T_{\rm ch}$ using the Saha equation approach and HRG in PCE with the \texttt{Thermal-FIST} package.
   The horizontal bands corresponds to the experimental data of the ALICE collaboration for most central Pb--Pb collisions~\cite{Adam:2015vda,Adam:2015yta,Acharya:2017bso}.
   The data point for $^3_\Lambda$H is reconstructed assuming a 25\% branching ratio of the $^3_\Lambda\text{H} \to ^3 \text{He} + \pi$ decay~\cite{Adam:2015yta}.
   The vertical yellow band in (a) corresponds to the kinetic freeze-out temperature $T_{\rm kin} = 113 \pm 12$~MeV extracted from blast-wave fits to the transverse momentum spectra of $\pi$, $\text{K}$, protons, d, and $^3$He~\cite{Adam:2015vda}. Taken from~\cite{Vovchenko:2019aoz}.}
\end{center}
\end{figure}

Often connections between the big bang or the early universe are drawn to compare to the the evolution and behaviour of the little bang observed in heavy-ion collisions (see for instance~\cite{Heinz:1999kb,Heinz:2013wva}. A new approach in that direction is the application of a (partial) chemical equilibrium picture on the yields of particles, allowing for a nuclear formation and disintegration as occuring in the early universe as indicated above via $p + n \leftrightarrow d + \gamma$ until photons decouple and the deuteron fraction is fixed. The corresponding rate equations are called Saha equation and in~\cite{Vovchenko:2019aoz} an equivalent equation is derived which is applicable for the little bang. It is shown that the light nuclei are less sensitive to disturbances by the cooling fireball (high pion density after chemical freeze-out) and the yield ratios of light (anti-)(hyper-)nuclei depend only smoothely on the actual temperature, as shown in Fig.~\ref{saha_2}.

In addition to the usual measured nuclei ratios also more exotic objects as strange dibaryons, depicted in Fig.~\ref{saha_2} have been calculated and the yield ratio of these is also rather constant as a function of the temperature. So a non-observation of these bound states might not be easily explained by the interaction with the fireball. 

\subsection{Criticism of the treatment of nuclei in a hadron resonance gas}
Recently, several ideas came up to overcome the "snowballs in hell" problem~\cite{BraunMunzinger:1994iq,BraunMunzinger:2001mh,Braun-Munzinger:2015cc} raised in the application of the thermal approach to the description of light (anti-)(hyper-)nuclei. In a recent work~\cite{Cai:2019jtk} the weakness of the thermal approach is demonstrated. In principle, one faces the issue that particles with binding energies of some MeV, or even only 130 keV for the (anti-)hypertriton, can be described at a temperature of 156 MeV. This large temperature and the many pions around would lead to the destruction of these loosely-bound objects. A way out could be the idea proposed in~\cite{Andronic:2017pug}, namely that the light (anti-)(hyper-)nuclei measured in the experiments are first formed as multi-quark states which only at a later stage become the object which is finally detected. In fact, the statistical hadronisation model is a model of hadronisation and not a model describing the cooling of the hadron resonance gas per se~\cite{BraunMunzinger:2003zz,Andronic:2017pug}. These facts are often used to favour the coalescence approach over the thermal approach. It is worth to mention that already in 1994 the yields of both model approaches have been found to be comparable~\cite{BraunMunzinger:1994iq}. Therefore, several ideas have been developed recently to find observables to discriminate between the two approaches.

One idea is to compare the production yields of $^{4}$Li and $^{4}$He with the predictions of the two models, as pointed out in~\cite{Mrowczynski:2016xqm,Bazak:2018hgl,Mrowczynski:2019yrr}. The difference in the size and the contribution from resonance decays should be significantly altered for the two approaches and thus the measured production yield should be different. The thermal model is mainly concerned about the mass and the quantum numbers of the produced state, whereas for more sophisticated coalescence models the size of the object, represented by the wave function of the nucleus, enters the calculation. The issue in this proposal is connected with the nature of
the $^{4}$Li, which is in a recent data collection~\cite{Tilley:1992zz} not a single resonance but rather a set of overlapping states. In connection to that information it is therefore not easy to get a single wave function for the object.

As pointed out for instance in~\cite{Bellini:2018epz,Sun:2018mqq} the difference can be strongly dependent on the size of the created fireball, so it would be beneficial to measure all accessible light (anti-)(hyper-)nuclei in as a function of multiplicity and compare the model predictions with the measurements. Since the $\Lambda$ separation energy of the hypertriton is so small, only 130 keV, this object is larger as a lead nucleus ($\sqrt{\langle r_{\mathrm{d}\Lambda}^2\rangle} = $ 10.6 fm~\cite{Braun-Munzinger:2018hat}) and would be the ultimate test case for this investigations. So the best way to discriminate thermal and coalescence models seems to be by comparing the measurments of different observables of ${}^{3}_{\Lambda}$H, $^{3}$He and $^{3}$H in different collision systems, which is planned in the upcoming runs of the LHC~\cite{Citron:2018lsq}.

\section{Summary and conclusion}
The description of the production of light (anti-)(hyper-)nuclei in a hadron resonance gas, or statistical-thermal model approach, has proven to be rather successful, despite the fact that the binding energies of these compound objects is small compared to the emitting fireball temperature. The grand-canonical ensemble is giving a very good description of the data at LHC, if light (anti-)nuclei are included in the description the particle yields covers nine orders of magnitude, by using only three parameters: chemical freeze-out temperature $T_{ch}$, baryo-chemical potential $\mu_B$ and the volume $V$.

To incorporate the tensions observed in this description several approaches have been investigated and discussed. A possible solution would be a sequential freeze-out scenario, where strangeness freezes out at a higher temperature and up and down at a lower temperature. On the other hand, this is in contradiction to the charm measurements at the LHC, which can be described by the statistical hadronisation model.

Repulsive forces can be modelled by van der Waals like forces or assuming an eigenvolume of the particles. If this eigenvolume is chosen to be different for the different nuclei the result is similar to the one when one size for all is assumed, whereas the extracted temperature might differ from the standard approach.

Another way is the treatment in a canonical ensemble, allowing for a (partial) equilibration of the system. This can cure the tension and at the same time be used to describe also smaller systems as pp or p--Pb collisions.
A partial chemical equilibration can also help understanding why the temperature extracted from grand-canonical thermal model fits is found to be in very good agreement with the data for light nuclei, despite the "snowballs in hell" problem.

Several ideas came up recently to finally pin down the true description of the production of light nuclei, so whether to favour the coalescence or the thermal model approach. The measurement of the (anti-)hypertriton as a function of the multiplicity in the collision seems to be a strong candidate as well as possible measurements of $^{4}$Li production, having the issue that the properties, i.e. the wave functions are not well known. So the measurement of the connected two-particle intensity correlations (often called femtoscopic measurements) might be more promising, to extract the production yield. Therefore, future measurements of different of the production of light (anti-)(hyper-)nuclei (stable, weakly decaying and unbound (anti-)nuclei) are needed to solve this outstanding issue.

\section*{Acknowledgements}
The author thanks A. Andronic, R. Bellwied, P. Braun-Munzinger, M. Floris, B. Hippolyte, A. Kalweit, M. Lorenz, the late H. Oeschler, M.~Petr\'{a}\v{n}, J. Rafelski, K. Redlich, J. Schukraft, J. Stachel, H. St\"{o}cker and V. Vovchenko for many useful discussions connected to the statistical-thermal model. This work was supported by BMBF through the FSP202 (F\"orderkennzeichen 05P15RFCA1).

\bibliographystyle{ws-ijmpe}
\bibliography{bib_nuclei_bd}

\begin{thebibliography}{100}

\bibitem{Cleymans:1992zc}
J.~Cleymans and H.~Satz, {\em Z. Phys.} {\bf C57}  (1993) 135,
  \href{http://arxiv.org/abs/hep-ph/9207204}{{\ttfamily arXiv:hep-ph/9207204
  [hep-ph]}}.

\bibitem{BraunMunzinger:1994xr}
P.~Braun-Munzinger, J.~Stachel, J.~Wessels and N.~Xu, {\em Phys. Lett. B} {\bf
  344}  (1995) 43, \href{http://arxiv.org/abs/nucl-th/9410026}{{\ttfamily
  arXiv:nucl-th/9410026 [nucl-th]}}.

\bibitem{BraunMunzinger:1995bp}
P.~Braun-Munzinger, J.~Stachel, J.~Wessels and N.~Xu, {\em Phys. Lett. B} {\bf
  365}  (1996) 1, \href{http://arxiv.org/abs/nucl-th/9508020}{{\ttfamily
  arXiv:nucl-th/9508020 [nucl-th]}}.

\bibitem{BraunMunzinger:1996mq}
P.~Braun-Munzinger and J.~Stachel, {\em Nucl. Phys.} {\bf A606}  (1996) 320,
  \href{http://arxiv.org/abs/nucl-th/9606017}{{\ttfamily arXiv:nucl-th/9606017
  [nucl-th]}}.

\bibitem{BraunMunzinger:1998cg}
P.~Braun-Munzinger and J.~Stachel, {\em Nucl. Phys. A} {\bf 638}  (1998) 3,
  \href{http://arxiv.org/abs/nucl-ex/9803015}{{\ttfamily arXiv:nucl-ex/9803015
  [nucl-ex]}}.

\bibitem{Becattini:2000jw}
F.~Becattini, J.~Cleymans, A.~Keranen, E.~Suhonen and K.~Redlich, {\em Phys.
  Rev.} {\bf C64}  (2001)   024901,
  \href{http://arxiv.org/abs/hep-ph/0002267}{{\ttfamily arXiv:hep-ph/0002267
  [hep-ph]}}.

\bibitem{Braun-Munzinger:2015hba}
P.~Braun-Munzinger, V.~Koch, T.~Sch\"afer and J.~Stachel, {\em Phys. Rept.}
  {\bf 621}  (2016) 76, \href{http://arxiv.org/abs/1510.00442}{{\ttfamily
  arXiv:1510.00442 [nucl-th]}}.

\bibitem{Mekjian:1977ei}
A.~Mekjian, {\em Phys. Rev. Lett.} {\bf 38}  (1977) 640.

\bibitem{Gosset:1988na}
J.~Gosset, J.~I. Kapusta and G.~D. Westfall, {\em Phys. Rev.} {\bf C18}  (1978)
  844.

\bibitem{Mekjian:1978us}
A.~Z. Mekjian, {\em Nucl. Phys.} {\bf A312}  (1978) 491.

\bibitem{Siemens:1979dz}
P.~J. Siemens and J.~I. Kapusta, {\em Phys. Rev. Lett.} {\bf 43}  (1979) 1486.

\bibitem{Stoecker:1981za}
H.~Stoecker, A.~A. Ogloblin and W.~Greiner, {\em Z. Phys.} {\bf A303}  (1981)
  259.

\bibitem{Hahn:1986mb}
D.~Hahn and H.~Stoecker, {\em Nucl. Phys.} {\bf A476}  (1988) 718.

\bibitem{Csernai:1986qf}
L.~P. Csernai and J.~I. Kapusta, {\em Phys. Rept.} {\bf 131}  (1986) 223.

\bibitem{BraunMunzinger:1994iq}
P.~Braun-Munzinger and J.~Stachel, {\em J. Phys.} {\bf G21}  (1995) L17,
  \href{http://arxiv.org/abs/nucl-th/9412035}{{\ttfamily arXiv:nucl-th/9412035
  [nucl-th]}}.

\bibitem{Andronic:2010qu}
A.~Andronic, P.~Braun-Munzinger, J.~Stachel and H.~Stocker, {\em Phys. Lett.}
  {\bf B697}  (2011) 203, \href{http://arxiv.org/abs/1010.2995}{{\ttfamily
  arXiv:1010.2995 [nucl-th]}}.

\bibitem{Steinheimer:2012tb}
J.~Steinheimer, K.~Gudima, A.~Botvina, I.~Mishustin, M.~Bleicher and
  H.~Stocker, {\em Phys. Lett.} {\bf B714}  (2012) 85,
  \href{http://arxiv.org/abs/1203.2547}{{\ttfamily arXiv:1203.2547 [nucl-th]}}.

\bibitem{Adam:2015vda}
 ALICE Collaboration (J.~Adam {\em et~al.}), {\em Phys. Rev.} {\bf C93}  (2016)
    024917, \href{http://arxiv.org/abs/1506.08951}{{\ttfamily arXiv:1506.08951
  [nucl-ex]}}.

\bibitem{Adam:2015yta}
 ALICE Collaboration (J.~Adam {\em et~al.}), {\em Phys. Lett.} {\bf B754}
  (2016) 360, \href{http://arxiv.org/abs/1506.08453}{{\ttfamily
  arXiv:1506.08453 [nucl-ex]}}.

\bibitem{Anticic:2016ckv}
 NA49 Collaboration (T.~Anticic {\em et~al.}), {\em Phys. Rev.} {\bf C94}
  (2016)   044906, \href{http://arxiv.org/abs/1606.04234}{{\ttfamily
  arXiv:1606.04234 [nucl-ex]}}.

\bibitem{Chen:2018tnh}
J.~Chen, D.~Keane, Y.~Ma, A.~Tang and Z.~Xu  (2018)
  \href{http://arxiv.org/abs/1808.09619}{{\ttfamily arXiv:1808.09619
  [nucl-ex]}}.

\bibitem{Braun-Munzinger:2018hat}
P.~Braun-Munzinger and B.~Dönigus, {\em Nucl. Phys.} {\bf A987}  (2019) 144,
  \href{http://arxiv.org/abs/1809.04681}{{\ttfamily arXiv:1809.04681
  [nucl-ex]}}.

\bibitem{Acharya:2017bso}
 ALICE Collaboration (S.~Acharya {\em et~al.}), {\em Nucl. Phys.} {\bf A971}
  (2018) 1, \href{http://arxiv.org/abs/1710.07531}{{\ttfamily arXiv:1710.07531
  [nucl-ex]}}.

\bibitem{Andronic:2017pug}
A.~Andronic, P.~Braun-Munzinger, K.~Redlich and J.~Stachel, {\em Nature} {\bf
  561}  (2018) 321, \href{http://arxiv.org/abs/1710.09425}{{\ttfamily
  arXiv:1710.09425 [nucl-th]}}.

\bibitem{hagedorn60}
R.~Hagedorn, {\em Phys. Rev. Lett.} {\bf 5} (Sep 1960) 276.

\bibitem{butler_pearson61}
S.~T. Butler and C.~A. Pearson, {\em Phys. Rev. Lett.} {\bf 7} (Jul 1961) 69.

\bibitem{butler_pearson63}
S.~T. Butler and C.~A. Pearson, {\em Phys. Rev.} {\bf 129} (Jan 1963) 836.

\bibitem{Hagedorn:1965st}
R.~Hagedorn, {\em Nuovo Cim. Suppl.} {\bf 3}  (1965) 147.

\bibitem{Hagedorn:1968zz}
R.~Hagedorn, {\em Nuovo Cim.} {\bf A56}  (1968) 1027.

\bibitem{Hagedorn:1984uy}
R.~Hagedorn and K.~Redlich, {\em Z. Phys.} {\bf C27}  (1985)   541.

\bibitem{BraunMunzinger:2001ip}
P.~Braun-Munzinger, D.~Magestro, K.~Redlich and J.~Stachel, {\em Phys. Lett. B}
  {\bf 518}  (2001) 41, \href{http://arxiv.org/abs/hep-ph/0105229}{{\ttfamily
  arXiv:hep-ph/0105229 [hep-ph]}}.

\bibitem{BraunMunzinger:2003zd}
P.~Braun-Munzinger, K.~Redlich and J.~Stachel  (2003)
  \href{http://arxiv.org/abs/nucl-th/0304013}{{\ttfamily arXiv:nucl-th/0304013
  [nucl-th]}}, Appeared in Quark Gluon Plasma 3, eds. R.C. Hwa and Xin-Nian
  Wang, World Scientific Publishing.

\bibitem{Becattini:2005xt}
F.~Becattini, J.~Manninen and M.~Gazdzicki, {\em Phys. Rev.} {\bf C73}  (2006)
   044905, \href{http://arxiv.org/abs/hep-ph/0511092}{{\ttfamily
  arXiv:hep-ph/0511092 [hep-ph]}}.

\bibitem{Andronic:2005yp}
A.~Andronic, P.~Braun-Munzinger and J.~Stachel, {\em Nucl. Phys.} {\bf A772}
  (2006) 167, \href{http://arxiv.org/abs/0511071}{{\ttfamily arXiv:0511071
  [nucl-th]}}.

\bibitem{Stachel:2013zma}
J.~Stachel, A.~Andronic, P.~Braun-Munzinger and K.~Redlich, {\em J. Phys. Conf.
  Ser.} {\bf 509}  (2014)   012019,
  \href{http://arxiv.org/abs/1311.4662}{{\ttfamily arXiv:1311.4662 [nucl-th]}}.

\bibitem{Becattini:2016xct}
F.~Becattini, J.~Steinheimer, R.~Stock and M.~Bleicher, {\em Phys. Lett.} {\bf
  B764}  (2017) 241, \href{http://arxiv.org/abs/1605.09694}{{\ttfamily
  arXiv:1605.09694 [nucl-th]}}.

\bibitem{Andronic:2018vqh}
A.~Andronic, P.~Braun-Munzinger, M.~K. Köhler and J.~Stachel, {\em Nucl.
  Phys.} {\bf A982}  (2019) 759,
  \href{http://arxiv.org/abs/1807.01236}{{\ttfamily arXiv:1807.01236
  [nucl-th]}}.

\bibitem{Stoecker:1984py}
H.~Stoecker, {\em J. Phys.} {\bf G10}  (1984) L111.

\bibitem{Hahn:1986pw}
D.~Hahn and H.~Stoecker, {\em Nucl. Phys. A} {\bf 452}  (1986) 723.

\bibitem{Vovchenko2019}
V.~Vovchenko, B.~D\"onigus, B.~Kardan, M.~Lorenz and H.~St\"ocker {, in
  preparation. (2020))}.

\bibitem{Shuryak:2019ikv}
E.~Shuryak and J.~M. Torres-Rincon  (2019)
  \href{http://arxiv.org/abs/1910.08119}{{\ttfamily arXiv:1910.08119
  [nucl-th]}}.

\bibitem{Cleymans:1998yb}
J.~Cleymans, H.~Oeschler and K.~Redlich, {\em Phys. Rev.} {\bf C59}  (1999)
  1663, \href{http://arxiv.org/abs/nucl-th/9809027}{{\ttfamily
  arXiv:nucl-th/9809027 [nucl-th]}}.

\bibitem{HADES2019}
 HADES Collaboration (J.~Adamczewski-Musch {\em et~al.}) {, in preparation.
  (2020))}.

\bibitem{thermalModel}
A.~Andronic, P.~Braun-Munzinger, J.~Stachel and H.~St\"{o}cker, {\em Phys.
  Lett.} {\bf B697}  (2011) 203,
  \href{http://arxiv.org/abs/1010.2995}{{\ttfamily arXiv:1010.2995 [nucl-th]}}.

\bibitem{pbm}
P.~Braun-Munzinger and J.~Stachel, {\em J. Phys.} {\bf G28}  (2002) 1971,
  \href{http://arxiv.org/abs/nucl-th/0112051}{{\ttfamily arXiv:nucl-th/0112051
  [nucl-th]}}.

\bibitem{pbm1}
P.~Braun-Munzinger and J.~Stachel, {\em J. Phys.} {\bf G21}  (1995) L17,
  \href{http://arxiv.org/abs/nucl-th/9412035}{{\ttfamily arXiv:nucl-th/9412035
  [nucl-th]}}.

\bibitem{anton_thermal}
A.~Andronic, P.~Braun-Munzinger and J.~Stachel, {\em Phys. Lett. B} {\bf 673}
  (2009)   142, \href{http://arxiv.org/abs/0812.1186}{{\ttfamily
  arXiv:0812.1186 [nucl-th]}}, Erratum ibid. 678 (2009) 516.

\bibitem{anton_sqm2016}
A.~Andronic, P.~Braun-Munzinger, K.~Redlich and J.~Stachel, { {Hadron yields,
  the chemical freeze-out and the QCD phase diagram}}, in {\em {Proceedings,
  16th International Conference on Strangeness in Quark Matter (SQM 2016):
  Berkeley, California, United States}\/},  {\em J. Phys. Conf. Ser.} {\bf 779}
  (2017), p. 012012.
\newblock \href{http://arxiv.org/abs/1611.01347}{{\ttfamily arXiv:1611.01347
  [nucl-th]}}.

\bibitem{Wheaton:2004qb}
S.~Wheaton and J.~Cleymans, {\em Comput. Phys. Commun.} {\bf 180}  (2009) 84,
  \href{http://arxiv.org/abs/hep-ph/0407174}{{\ttfamily arXiv:hep-ph/0407174
  [hep-ph]}}.

\bibitem{Wheaton:2004vg}
S.~Wheaton and J.~Cleymans, {\em J. Phys. G} {\bf G31}  (2005) S1069,
  \href{http://arxiv.org/abs/hep-ph/0412031}{{\ttfamily arXiv:hep-ph/0412031
  [hep-ph]}}.

\bibitem{rafelski0}
G.~Torrieri, S.~Steinke, W.~Broniowski, W.~Florkowski, J.~Letessier and
  J.~Rafelski, {\em Comput. Phys. Commun.} {\bf 167}  (2005)   229.

\bibitem{rafelski1}
G.~Torrieri, S.~Jeon, J.~Letessier and J.~Rafelski, {\em Comput. Phys. Commun.}
  {\bf 175}  (2006)   635.

\bibitem{Torrieri:2004zz}
G.~Torrieri, S.~Steinke, W.~Broniowski, W.~Florkowski, J.~Letessier {\em
  et~al.}, {\em Comput. Phys. Commun.} {\bf 167}  (2005) 229,
  \href{http://arxiv.org/abs/nucl-th/0404083}{{\ttfamily arXiv:nucl-th/0404083
  [nucl-th]}}.

\bibitem{ThermalFIST}
The \texttt{Thermal-FIST} package,
  \url{https://github.com/vlvovch/Thermal-FIST}.

\bibitem{Vovchenko:2019pjl}
V.~Vovchenko and H.~Stoecker, {\em Comput. Phys. Commun.} {\bf 244}  (2019)
  295, \href{http://arxiv.org/abs/1901.05249}{{\ttfamily arXiv:1901.05249
  [nucl-th]}}.

\bibitem{Floris2014103}
M.~Floris, {\em Nuclear Physics A} {\bf 931}  (2014) 103 .

\bibitem{Nagamiya:1984vk}
S.~Nagamiya, J.~Randrup and T.~J.~M. Symons, {\em Ann. Rev. Nucl. Part. Sci.}
  {\bf 34}  (1984) 155.

\bibitem{Fachini:2006ch}
P.~Fachini, {\em AIP Conf. Proc.} {\bf 857}  (2006) 62,
  \href{http://arxiv.org/abs/hep-ex/0605102}{{\ttfamily arXiv:hep-ex/0605102
  [hep-ex]}}.

\bibitem{Bellwied:2013cta}
R.~Bellwied, S.~Borsanyi, Z.~Fodor, S.~D. Katz and C.~Ratti, {\em Phys. Rev.
  Lett.} {\bf 111}  (2013)   202302,
  \href{http://arxiv.org/abs/1305.6297}{{\ttfamily arXiv:1305.6297 [hep-lat]}}.

\bibitem{Alba:2015iva}
P.~Alba, R.~Bellwied, M.~Bluhm, V.~Mantovani~Sarti, M.~Nahrgang and C.~Ratti,
  {\em Phys. Rev.} {\bf C92}  (2015)   064910,
  \href{http://arxiv.org/abs/1504.03262}{{\ttfamily arXiv:1504.03262
  [hep-ph]}}.

\bibitem{Bellwied:2017uat}
R.~Bellwied  (2017) \href{http://arxiv.org/abs/1711.00514}{{\ttfamily
  arXiv:1711.00514 [nucl-ex]}}, [EPJ Web Conf.17,10200(2018)].

\bibitem{Ratti:2019ytu}
C.~Ratti, R.~Bellwied, J.~Noronha-Hostler, P.~Parotto, I.~Portillo~Vazquez and
  J.~M. Stafford, {\em Nucl. Phys.} {\bf A982}  (2019) 799.

\bibitem{Chatterjee:2013yga}
S.~Chatterjee, R.~M. Godbole and S.~Gupta, {\em Phys. Lett.} {\bf B727}  (2013)
  554, \href{http://arxiv.org/abs/1306.2006}{{\ttfamily arXiv:1306.2006
  [nucl-th]}}.

\bibitem{Chatterjee:2014ysa}
S.~Chatterjee and B.~Mohanty, {\em Phys. Rev.} {\bf C90}  (2014)   034908,
  \href{http://arxiv.org/abs/1405.2632}{{\ttfamily arXiv:1405.2632 [nucl-th]}}.

\bibitem{Chatterjee:2015fua}
S.~Chatterjee, S.~Das, L.~Kumar, D.~Mishra, B.~Mohanty, R.~Sahoo and N.~Sharma,
  {\em Adv. High Energy Phys.} {\bf 2015}  (2015)   349013.

\bibitem{Chatterjee:2016dld}
S.~Chatterjee and B.~Mohanty, {\em Springer Proc. Phys.} {\bf 174}  (2016) 165.

\bibitem{star}
 STAR Collaboration (B.~I. Abelev {\em et~al.}), {\em Science} {\bf 328}
  (2010)  ~58, \href{http://arxiv.org/abs/1003.2030}{{\ttfamily arXiv:1003.2030
  [nucl-ex]}}.

\bibitem{Andronic:2019wva}
A.~Andronic, P.~Braun-Munzinger, M.~K. Köhler, K.~Redlich and J.~Stachel, {\em
  Phys. Lett.} {\bf B797}  (2019)   134836,
  \href{http://arxiv.org/abs/1901.09200}{{\ttfamily arXiv:1901.09200
  [nucl-th]}}.

\bibitem{hypertriton}
 ALICE Collaboration (J.~Adam {\em et~al.}), {\em Phys. Lett. B} {\bf 754}
  (2016) 360, \href{http://arxiv.org/abs/1506.08453}{{\ttfamily
  arXiv:1506.08453 [nucl-ex]}}.

\bibitem{Hagedorn:1980kb}
R.~Hagedorn and J.~Rafelski, {\em Phys. Lett.} {\bf 97B}  (1980)   136.

\bibitem{Rischke:1991ke}
D.~H. Rischke, M.~I. Gorenstein, H.~Stoecker and W.~Greiner, {\em Z. Phys.}
  {\bf C51}  (1991) 485.

\bibitem{Vovchenko:2015cbk}
V.~Vovchenko and H.~Stöcker, {\em J. Phys.} {\bf G44}  (2017)   055103,
  \href{http://arxiv.org/abs/1512.08046}{{\ttfamily arXiv:1512.08046
  [hep-ph]}}.

\bibitem{Vovchenko:2016ebv}
V.~Vovchenko and H.~Stoecker, {\em Phys. Rev.} {\bf C95}  (2017)   044904,
  \href{http://arxiv.org/abs/1606.06218}{{\ttfamily arXiv:1606.06218
  [hep-ph]}}.

\bibitem{Vovchenko:2016eby}
V.~Vovchenko, M.~I. Gorenstein, L.~M. Satarov and H.~Stöcker, { {Chemical
  freeze-out conditions in hadron resonance gas}}, in {\em {Proceedings,
  International Symposium on New Horizons in Fundamental Physics: From Neutrons
  Nuclei via Superheavy Elements and Supercritical Fields to Neutron Stars and
  Cosmic Rays: Makutsi, South Africa, November 23-29, 2015}\/},  (2017), pp.
  127--137.
\newblock \href{http://arxiv.org/abs/1606.06350}{{\ttfamily arXiv:1606.06350
  [hep-ph]}}.

\bibitem{Vovchenko:2016mwg}
V.~Vovchenko and H.~Stoecker, {\em J. Phys. Conf. Ser.} {\bf 779}  (2017)
  012078, \href{http://arxiv.org/abs/1610.02346}{{\ttfamily arXiv:1610.02346
  [nucl-th]}}.

\bibitem{PhysRevC.95.024902}
L.~M. Satarov, V.~Vovchenko, P.~Alba, M.~I. Gorenstein and H.~Stoecker, {\em
  Phys. Rev. C} {\bf 95} (Feb 2017)   024902.

\bibitem{Vovchenko:2017zpj}
V.~Vovchenko, A.~Motornenko, P.~Alba, M.~I. Gorenstein, L.~M. Satarov and
  H.~Stoecker, {\em Phys. Rev.} {\bf C96}  (2017)   045202,
  \href{http://arxiv.org/abs/1707.09215}{{\ttfamily arXiv:1707.09215
  [nucl-th]}}.

\bibitem{Vovchenko:2016rkn}
V.~Vovchenko, M.~I. Gorenstein and H.~Stoecker, {\em Phys. Rev. Lett.} {\bf
  118}  (2017)   182301, \href{http://arxiv.org/abs/1609.03975}{{\ttfamily
  arXiv:1609.03975 [hep-ph]}}.

\bibitem{Alba:2016hwx}
P.~Alba, V.~Vovchenko, M.~I. Gorenstein and H.~Stoecker, {\em Nucl. Phys.} {\bf
  A974}  (2018) 22, \href{http://arxiv.org/abs/1606.06542}{{\ttfamily
  arXiv:1606.06542 [hep-ph]}}.

\bibitem{Vovchenko:2017drx}
V.~Vovchenko, A.~Motornenko, M.~I. Gorenstein and H.~Stoecker, {\em Phys. Rev.}
  {\bf C97}  (2018)   035202, \href{http://arxiv.org/abs/1710.00693}{{\ttfamily
  arXiv:1710.00693 [nucl-th]}}.

\bibitem{Vovchenko:2017ygz}
V.~Vovchenko, M.~I. Gorenstein and H.~Stoecker, {\em Eur. Phys. J.} {\bf A54}
  (2018)  ~16, \href{http://arxiv.org/abs/1709.10097}{{\ttfamily
  arXiv:1709.10097 [nucl-th]}}.

\bibitem{Poberezhnyuk:2017yhx}
R.~V. Poberezhnyuk, V.~Vovchenko, D.~V. Anchishkin and M.~I. Gorenstein, {\em
  Int. J. Mod. Phys.} {\bf E26}  (2017)   1750061,
  \href{http://arxiv.org/abs/1708.05605}{{\ttfamily arXiv:1708.05605
  [nucl-th]}}.

\bibitem{Andronic:2018qqt}
A.~Andronic, P.~Braun-Munzinger, B.~Friman, P.~M. Lo, K.~Redlich and
  J.~Stachel, {\em Phys. Lett.} {\bf B792}  (2019) 304,
  \href{http://arxiv.org/abs/1808.03102}{{\ttfamily arXiv:1808.03102
  [hep-ph]}}.

\bibitem{Rafelski:1980gk}
J.~Rafelski and M.~Danos, {\em Phys. Lett.} {\bf 97B}  (1980) 279.

\bibitem{Cleymans:1990mn}
J.~Cleymans, K.~Redlich and E.~Suhonen, {\em Z. Phys.} {\bf C51}  (1991) 137.

\bibitem{Becattini:1995if}
F.~Becattini, {\em Z. Phys.} {\bf C69}  (1996) 485.

\bibitem{Andronic:2008ev}
A.~Andronic, F.~Beutler, P.~Braun-Munzinger, K.~Redlich and J.~Stachel, {\em
  Phys. Lett.} {\bf B675}  (2009) 312,
  \href{http://arxiv.org/abs/0804.4132}{{\ttfamily arXiv:0804.4132 [hep-ph]}}.

\bibitem{Becattini:2008tx}
F.~Becattini, P.~Castorina, J.~Manninen and H.~Satz, {\em Eur. Phys. J.} {\bf
  C56}  (2008) 493, \href{http://arxiv.org/abs/0805.0964}{{\ttfamily
  arXiv:0805.0964 [hep-ph]}}.

\bibitem{Becattini:1997rv}
F.~Becattini and U.~W. Heinz, {\em Z. Phys.} {\bf C76}  (1997) 269,
  \href{http://arxiv.org/abs/hep-ph/9702274}{{\ttfamily arXiv:hep-ph/9702274
  [hep-ph]}}, [Erratum: Z. Phys.C76,578(1997)].

\bibitem{Becattini:2010sk}
F.~Becattini, P.~Castorina, A.~Milov and H.~Satz, {\em Eur. Phys. J.} {\bf C66}
   (2010) 377, \href{http://arxiv.org/abs/0911.3026}{{\ttfamily arXiv:0911.3026
  [hep-ph]}}.

\bibitem{Kraus:2008fh}
I.~Kraus, J.~Cleymans, H.~Oeschler and K.~Redlich, {\em Phys. Rev.} {\bf C79}
  (2009)   014901, \href{http://arxiv.org/abs/0808.0611}{{\ttfamily
  arXiv:0808.0611 [hep-ph]}}.

\bibitem{Adam:2015vsf}
 ALICE Collaboration (J.~Adam {\em et~al.}), {\em Phys. Lett.} {\bf B758}
  (2016) 389, \href{http://arxiv.org/abs/1512.07227}{{\ttfamily
  arXiv:1512.07227 [nucl-ex]}}.

\bibitem{Vislavicius:2016rwi}
V.~Vislavicius and A.~Kalweit  (2016)
  \href{http://arxiv.org/abs/1610.03001}{{\ttfamily arXiv:1610.03001
  [nucl-ex]}}.

\bibitem{Vovchenko:2018fiy}
V.~Vovchenko, B.~D\"onigus and H.~St\"ocker, {\em Phys. Lett.} {\bf B785}
  (2018) 171, \href{http://arxiv.org/abs/1808.05245}{{\ttfamily
  arXiv:1808.05245 [hep-ph]}}.

\bibitem{Castorina:2013mba}
P.~Castorina and H.~Satz, {\em Int. J. Mod. Phys.} {\bf E23}  (2014)   1450019,
  \href{http://arxiv.org/abs/1310.6932}{{\ttfamily arXiv:1310.6932 [hep-ph]}}.

\bibitem{Acharya:2017fvb}
 ALICE Collaboration (S.~Acharya {\em et~al.}), {\em Phys. Rev.} {\bf C97}
  (2018)   024615, \href{http://arxiv.org/abs/1709.08522}{{\ttfamily
  arXiv:1709.08522 [nucl-ex]}}.

\bibitem{Heinz:1999kb}
U.~W. Heinz, {\em Nucl. Phys.} {\bf A661}  (1999) 140,
  \href{http://arxiv.org/abs/nucl-th/9907060}{{\ttfamily arXiv:nucl-th/9907060
  [nucl-th]}}.

\bibitem{Heinz:2013wva}
U.~W. Heinz, {\em J. Phys. Conf. Ser.} {\bf 455}  (2013)   012044,
  \href{http://arxiv.org/abs/1304.3634}{{\ttfamily arXiv:1304.3634 [nucl-th]}}.

\bibitem{Vovchenko:2019aoz}
V.~Vovchenko, K.~Gallmeister, J.~Schaffner-Bielich and C.~Greiner  (2019)
  \href{http://arxiv.org/abs/1903.10024}{{\ttfamily arXiv:1903.10024
  [hep-ph]}}.

\bibitem{BraunMunzinger:2001mh}
P.~Braun-Munzinger and J.~Stachel, {\em J. Phys. G} {\bf 28}  (2002) 1971,
  \href{http://arxiv.org/abs/nucl-th/0112051}{{\ttfamily arXiv:nucl-th/0112051
  [nucl-th]}}.

\bibitem{Braun-Munzinger:2015cc}
P.~Braun-Munzinger, B.~Dönigus and N.~Löher, {\em CERN Courier} {\bf
  September}  (2015)  ~26.

\bibitem{Cai:2019jtk}
Y.~Cai, T.~D. Cohen, B.~A. Gelman and Y.~Yamauchi, {\em Phys. Rev.} {\bf C100}
  (2019)   024911, \href{http://arxiv.org/abs/1905.02753}{{\ttfamily
  arXiv:1905.02753 [nucl-th]}}.

\bibitem{BraunMunzinger:2003zz}
P.~Braun-Munzinger, J.~Stachel and C.~Wetterich, {\em Phys. Lett.} {\bf B596}
  (2004) 61, \href{http://arxiv.org/abs/nucl-th/0311005}{{\ttfamily
  arXiv:nucl-th/0311005 [nucl-th]}}.

\bibitem{Mrowczynski:2016xqm}
S.~Mrowczynski, {\em Acta Phys. Polon.} {\bf B48}  (2017)   707,
  \href{http://arxiv.org/abs/1607.02267}{{\ttfamily arXiv:1607.02267
  [nucl-th]}}.

\bibitem{Bazak:2018hgl}
S.~Bazak and S.~Mrowczynski, {\em Mod. Phys. Lett.} {\bf A33}  (2018)
  1850142, \href{http://arxiv.org/abs/1802.08212}{{\ttfamily arXiv:1802.08212
  [nucl-th]}}.

\bibitem{Mrowczynski:2019yrr}
S.~Mrowczynski and P.~Slon  (2019)
  \href{http://arxiv.org/abs/1904.08320}{{\ttfamily arXiv:1904.08320
  [nucl-th]}}.

\bibitem{Tilley:1992zz}
D.~R. Tilley, H.~R. Weller and G.~M. Hale, {\em Nucl. Phys.} {\bf A541}  (1992)
  1.

\bibitem{Bellini:2018epz}
F.~Bellini and A.~P. Kalweit, {\em Phys. Rev.} {\bf C99}  (2019)   054905,
  \href{http://arxiv.org/abs/1807.05894}{{\ttfamily arXiv:1807.05894
  [hep-ph]}}.

\bibitem{Sun:2018mqq}
K.-J. Sun, C.~M. Ko and B.~Dönigus, {\em Phys. Lett.} {\bf B792}  (2019) 132,
  \href{http://arxiv.org/abs/1812.05175}{{\ttfamily arXiv:1812.05175
  [nucl-th]}}.

\bibitem{Citron:2018lsq}
Z.~Citron {\em et~al.}  (2019)
  \href{http://arxiv.org/abs/1812.06772}{{\ttfamily arXiv:1812.06772
  [hep-ph]}}, CERN-LPCC-2018-07.

\end{thebibliography}

\end{document}